\documentclass{aa}

\usepackage{txfonts}
\usepackage[breaklinks=false]{hyperref}
\usepackage{color}
\usepackage[dvipsnames]{xcolor}

\newcommand{\lcdm}{$\Lambda$CDM}
\newcommand{\lya}{Ly$\alpha$}
\newcommand{\lyb}{Ly$\beta$}

\newcommand{\apar}{\alpha_{\parallel}}
\newcommand{\aperp}{\alpha_{\perp}}
\newcommand{\rpar}{r_{\parallel}}
\newcommand{\rperp}{r_{\perp}}
\newcommand{\chiSquareMin}{\chi^{2}_{\rm min}}
\newcommand{\LyaForest}{Ly$\alpha$~forest~}
\newcommand{\hMpc}{h^{-1}~\mathrm{Mpc}}
\newcommand{\DHh}{D_{H}}
\newcommand{\DMm}{D_{M}}

\DeclareMathOperator*{\sinc}{sinc}

\makeatletter
\renewcommand*\aa@pageof{, page \thepage{} of \pageref*{LastPage}}
\makeatother

\begin{document}

\title{Baryon acoustic oscillations from the cross-correlation of Ly$\alpha$ absorption and quasars in eBOSS DR14}

\author{
Michael~Blomqvist\inst{1}\thanks{E-mail: michael.blomqvist@lam.fr},
H{\'e}lion~du~Mas~des~Bourboux\inst{2},
Nicol{\'a}s~G.~Busca\inst{3},
Victoria~de~Sainte~Agathe\inst{3},
James~Rich\inst{4},
Christophe~Balland\inst{3},
Julian~E.~Bautista\inst{5},
Kyle~Dawson\inst{2},
Andreu~Font-Ribera\inst{6},
Julien~Guy\inst{7},
Jean-Marc~Le Goff\inst{4},
Nathalie~Palanque-Delabrouille\inst{4},
Will~J.~Percival\inst{8},
Ignasi~P{\'e}rez-R{\`a}fols\inst{1},
Matthew~M.~Pieri\inst{1},
Donald~P.~Schneider\inst{9,10},
An\v{z}e~Slosar\inst{11},
Christophe~Y{\`e}che\inst{4}
}

\institute{
Aix Marseille Univ, CNRS, CNES, LAM, Marseille, France
\and
Department of Physics and Astronomy, University of Utah, 115 S. 1400 E., Salt Lake City, UT 84112, U.S.A.
\and
Sorbonne Universit{\'e}, CNRS/IN2P3, Laboratoire de Physique Nucl{\'e}aire et de Hautes Energies, LPNHE, 4 Place Jussieu, F-75252 Paris, France
\and
IRFU, CEA, Universit{\'e} Paris-Saclay, F-91191 Gif-sur-Yvette, France
\and
Institute of Cosmology \& Gravitation, University of Portsmouth, Dennis Sciama Building, Portsmouth, PO1 3FX, UK
\and
University College London, Gower St, Kings Cross, London WC1E 6BT
\and
Lawrence Berkeley National Laboratory, 1 Cyclotron Road, Berkeley, CA 94720, U.S.A.
\and
Department of Physics and Astronomy, University of Waterloo, 200 University Ave. W., Waterloo ON N2L 3G1, Canada
\and
Department of Astronomy and Astrophysics, The Pennsylvania State University, University Park, PA 16802, U.S.A.
\and
Institute for Gravitation and the Cosmos, The Pennsylvania State University, University Park, PA 16802, U.S.A.
\and
Brookhaven National Laboratory, 2 Center Road,  Upton, NY 11973, U.S.A.
}

\abstract{We present a measurement of the baryon acoustic oscillation (BAO) scale at redshift $z=2.35$ from the three-dimensional correlation of Lyman-$\alpha$ (Ly$\alpha$) forest absorption and quasars. The study uses 266,590 quasars in the redshift range $1.77<z<3.5$ from the Sloan Digital Sky Survey (SDSS) Data Release 14 (DR14). The sample includes the first two years of observations by the SDSS-IV extended Baryon Oscillation Spectroscopic Survey (eBOSS), providing new quasars and re-observations of BOSS quasars for improved statistical precision. Statistics are further improved by including Ly$\alpha$ absorption occurring in the Ly$\beta$ wavelength band of the spectra. From the measured BAO peak position along and across the line of sight, we determined the Hubble distance $\DHh$ and the comoving angular diameter distance $\DMm$ relative to the sound horizon at the drag epoch $r_{d}$: $\DHh(z=2.35)/r_{d}=9.20\pm 0.36$ and $\DMm(z=2.35)/r_{d}=36.3\pm 1.8$. These results are consistent at $1.5\sigma$ with the prediction of the best-fit spatially-flat cosmological model with the cosmological constant reported for the Planck (2016) analysis of cosmic microwave background anisotropies. Combined with the Ly$\alpha$ auto-correlation measurement presented in a companion paper, the BAO measurements at $z=2.34$ are within $1.7\sigma$ of the predictions of this model.}


\authorrunning{M. Blomqvist et al.}
\titlerunning{BAO from the Ly$\alpha$-quasar cross-correlation of eBOSS DR14}
\maketitle

%===================================================
\section{Introduction}
%===================================================

The baryon acoustic oscillation (BAO) peak in the cosmological matter correlation function at a distance corresponding to the sound horizon, $r_d\sim100\hMpc$, has been seen at several redshifts using a variety of tracers. Following the original measurements \citep{2005ApJ...633..560E,2005MNRAS.362..505C}, the most precise results have been obtained using bright galaxies in the redshift range  $0.35<z<0.65$ \citep{2012MNRAS.427.3435A,2014MNRAS.439...83A,2014MNRAS.441...24A,2017MNRAS.470.2617A} from the Baryon Oscillation Spectroscopy Survey (BOSS; \citealt{2013AJ....145...10D}) of the Sloan Digital Sky Survey-III (SDSS-III; \citealt{2011AJ....142...72E}). Other measurements using galaxies cover the range $0.1<z<0.8$ \citep{2007MNRAS.381.1053P,2010MNRAS.401.2148P,2011MNRAS.416.3017B,2011MNRAS.415.2892B,2012MNRAS.427.2132P,2012MNRAS.427.2168M,2012MNRAS.426..226C,2013MNRAS.431.2834X,2015MNRAS.449..835R,2018ApJ...863..110B}. At higher redshift, the peak has been seen in the correlation function of quasars at a mean redshift $z\sim1.5$ \citep{2018MNRAS.473.4773A,GilMarin18,Hou18,Zarrouk18}
and in the flux-transmission correlation function in Lyman-$\alpha$ (Ly$\alpha$) forests at $z\sim 2.3$ \citep{2013A&A...552A..96B,2013JCAP...04..026S,2013JCAP...03..024K,2015A&A...574A..59D,2017A&A...603A..12B} and in the forest cross-correlation with quasars \citep{2014JCAP...05..027F,2017A&A...608A.130D}. These observations all yield measurements of comoving angular-diameter distances and Hubble distances at the corresponding redshift, $\DMm(z)/r_d$ and $\DHh(z)/r_d=c/(H(z)r_d)$, relative to the sound horizon.

BAO measurements have found an important role in testing the robustness of the spatially-flat cosmology with cold dark matter and the cosmological constant (\lcdm) that is consistent with observed cosmic microwave background (CMB) anisotropies \citep{2016A&A...594A..13P}. While the parameters of this model are precisely determined by the CMB data by itself, more general models are not constrained as well. Most significantly, adding BAO data improves constraints on curvature \citep{2016A&A...594A..13P}. The addition of BAO and type Ia supernova (SN~Ia) data \citep{2014A&A...568A..22B} generalizes the ``CMB'' measurement of $H_0$, which assumes flatness, to give an ``inverse-ladder'' measurement of $H_0$ \citep{2015PhRvD..92l3516A} that can be compared with distance-ladder measurements \citep{2016ApJ...826...56R,2018ApJ...855..136R,2018ApJ...861..126R}. Here, the inverse-ladder method uses the CMB-determined value of $r_d$ to define BAO-determined absolute distances to intermediate redshifts, $z\sim0.5$, which can then be used to calibrate SN~Ia luminosities. The usual distance ladder calibrates the SN~Ia luminosity using Cepheid luminosities, themselves calibrated through geometrical distance determinations.

A third use of BAO data is to determine \lcdm~parameters in a CMB-independent way. The \LyaForest auto- and cross-correlations that BOSS has pioneered are critical when gathering such measurements. It is striking that the o$\Lambda$CDM parameters $(\Omega_M,\Omega_\Lambda)$ determined by this method are in good agreement with the CMB values determined by assuming flat \lcdm~\citep{2015PhRvD..92l3516A}.

The individual BAO measurements of $\DMm(z)/r_d$ and $\DHh(z)/r_d$ are generally in good agreement with the CMB flat \lcdm~model. The largest single discrepancy, 1.8 standard deviations, is that of the BOSS (SDSS Data Release 12) measurement of the \lya~forest-quasar cross-correlation of \citet{2017A&A...608A.130D} (hereafter dMdB17). In this paper, we update this analysis with new quasars and forests from the SDSS Data Release 14 (DR14; \citealt{2018ApJS..235...42A,2018A&A...613A..51P}) obtained in the extended Baryon Oscillation Spectroscopy Survey (eBOSS) program \citep{2016AJ....151...44D} of SDSS-IV \citep{2017AJ....154...28B}. This data set has been previously used to measure the cross-correlation between quasars and the flux in the ``CIV forest'' due to absorption by triply-ionized carbon \citep{2018JCAP...05..029B}.

Besides the addition of new quasars and forests, our analysis differs in a few ways with that of dMdB17. Most importantly, we expand the wavelength range of the forest from the nominal \lya~forest, $104.0<\lambda_{\rm rf}<120.0~\mathrm{nm}$, to include \lya~absorption ($\lambda_{\alpha}=121.567$~nm) in the \lyb~region of the spectra, $97.4<\lambda_{\rm rf}<102.0~\mathrm{nm}$, thus increasing the statistical power of the sample. The procedure for fitting the correlation function is also slightly modified by including relativistic corrections \citep{2014PhRvD..89h3535B,2016JCAP...02..051I}. Furthermore, we divide the data to report BAO measurements for two redshift bins. We have not developed new sets of mock spectra beyond those used in dMdB17. We refer to Section 6 of dMdB17 for the analysis of those mocks and the tests used to justify the analysis procedure.

The organization of this paper follows closely that of dMdB17.
Section~\ref{section::data} describes the DR14 data set used in this study. Section~\ref{section::Measurement_of_the_transmission_field} summarizes the measurement of the flux-transmission field.
Section~\ref{section::The_Lya_forest_quasar_cross_correlation}
describes the measurement of the cross-correlation of the transmission field with quasars and the associated covariance matrix.
We also derive the ``distortion matrix'' that describes how the measured cross-correlation is related to the underlying physical cross-correlation. Section~\ref{section::model} describes our theoretical model of the cross-correlation. Section~\ref{section::fits} presents the fits to the observed correlation function and section~\ref{section::combwithauto} combines these results with those from the \lya~auto-correlation function presented in a companion paper \citep{2019agathe}. Section~\ref{section::cosmology} summarizes the constraints on cosmological parameters derived from these results and those from \citet{2019agathe}. Our conclusions are presented in Section~\ref{section::conclusions}. The measurements presented in this paper were made using the publicly available Python package \texttt{picca}\footnote{Package for Igm Cosmological-Correlations Analyses (\texttt{picca}) is available at \url{https://github.com/igmhub/picca/}} developed by our team.

%===================================================
\section{Data sample and reduction}
\label{section::data}
%===================================================

\begin{figure*}[tb]
\centering
\includegraphics[width=0.8\textwidth]{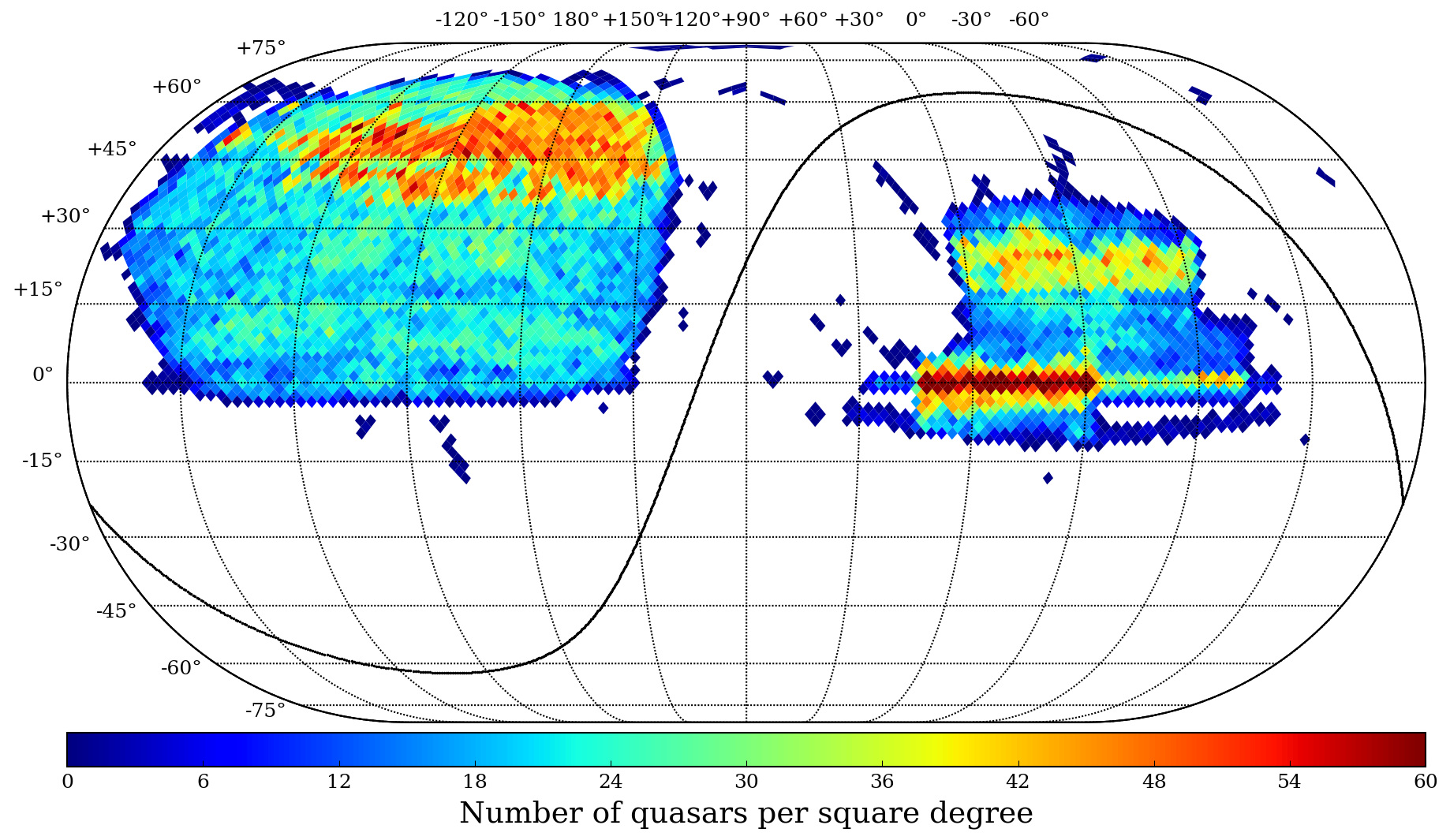}
\caption{Sky distribution for sample of 266,590 tracer quasars ($1.77 < z_{\rm q} < 3.5$) from DR14Q in J2000 equatorial coordinates. The solid black curve is the Galactic plane. The high-density regions are the eBOSS and SEQUELS observations (for the northern regions of the two Galactic hemispheres) and SDSS-stripe 82 (for declination $\delta\sim0$). The discontiguous small areas contain only SDSS DR7 quasars.}
\label{figure::sky}
\end{figure*}

\begin{figure}[tb]
\centering
\includegraphics[width=\columnwidth]{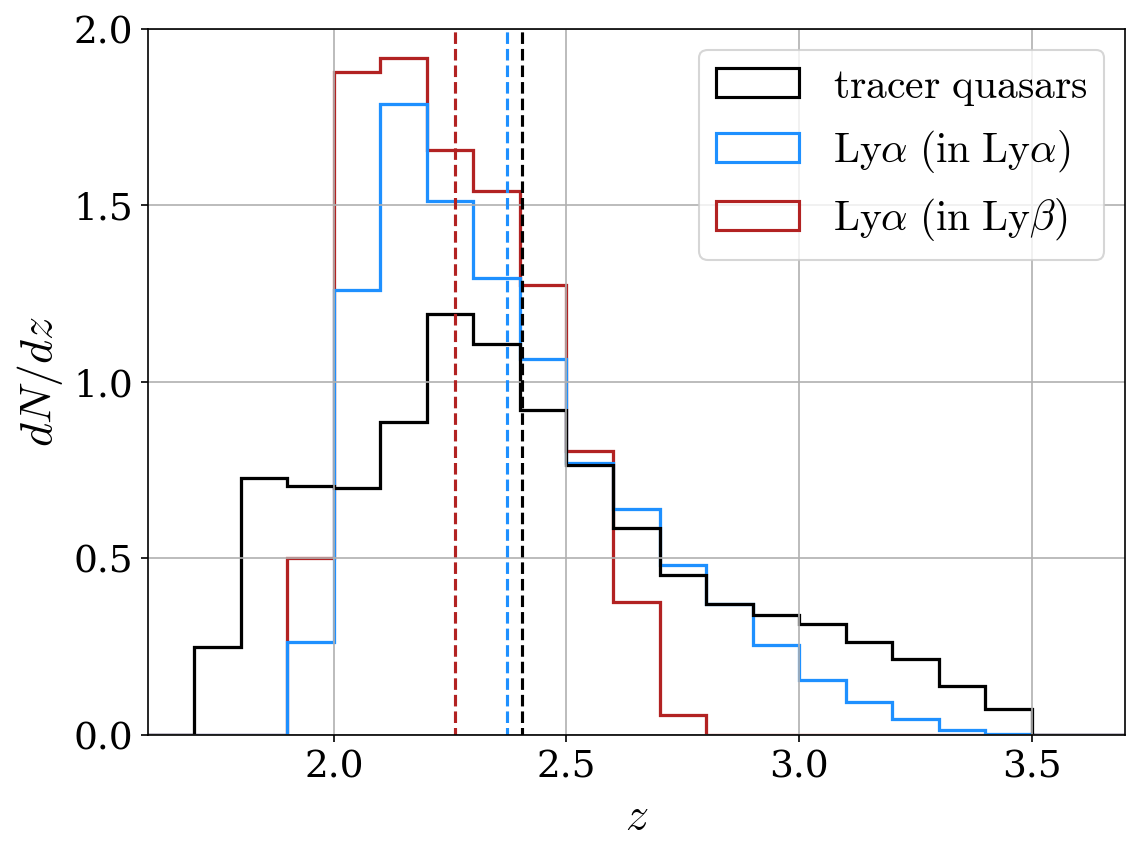}
\caption{Normalized redshift distributions for tracer quasars (black) and Ly$\alpha$ forest absorption pixels of Ly$\alpha$ region (blue) and Ly$\beta$ region (red). The histograms include 266,590 tracer quasars, \mbox{$30.2\times10^{6}$} pixels in the Ly$\alpha$ region, and $4.0\times10^{6}$ pixels in the Ly$\beta$ region. The vertical dashed lines show the mean value of each distribution: $\overline{z}= 2.40$ (tracer quasars), 2.37 (in Ly$\alpha$), 2.26 (in Ly$\beta$).}
\label{figure::zhistosamples}
\end{figure}

The quasars and forests used in this study are drawn from SDSS DR14. This release includes data from DR12 taken in the first two generations SDSS-I/II, in the BOSS program of SDSS-III and in the eBOSS pilot program SEQUELS \citep{2015ApJS..221...27M}. These data were used in the measurement of the quasar-forest
cross-correlation of dMdB17. Here, we use in addition data from the first two years of the eBOSS program and the completed SEQUELS.

The quasar target selection for BOSS, summarized in \citet{2012ApJS..199....3R}, combines different targeting methods described in \citet{2010A&A...523A..14Y}, \citet{2011ApJ...743..125K}, and \citet{2011ApJ...729..141B}. The methods employed for eBOSS quasar target selection are described in \citet{2015ApJS..221...27M} and \citet{2016A&A...587A..41P}.

The catalog of identified quasars, DR14Q \citep{2018A&A...613A..51P}, includes 266,590 quasars\footnote{Excludes plate 7235 for which object identification \texttt{\detokenize{THING_ID=}}0.} in the redshift range $1.77<z_{\rm q}<3.5$. The distribution on the sky of these quasars is shown in Fig. \ref{figure::sky} and the redshift distribution in Fig. \ref{figure::zhistosamples}.

All spectra used for this analysis were obtained using the BOSS spectrograph \citep{2013AJ....146...32S} on the 2.5 m SDSS telescope \citep{2006AJ....131.2332G} at Apache Point Observatory (APO). The spectrograph covers observed wavelengths \mbox{$360.0\lesssim\lambda\lesssim1040.0~{\rm nm}$}, with a resolving power $R\equiv \lambda/\Delta\lambda_{\rm FWHM}$ increasing from $\sim1300$ to $\sim2600$ across the wavelength range. The data were processed by the eBOSS pipeline, the same (but a marginally updated version) as that used for the cross-correlation measurement of dMdB17. The pipeline performs wavelength calibration, flux calibration and sky subtraction of the spectra. The individual exposures (typically four of 15 minutes) of a given object are combined into a coadded spectrum that is rebinned onto pixels on a uniform grid with $\Delta\log_{10}(\lambda)=10^{-4}$ (velocity width $\Delta v\approx 69$~km~s$^{-1}$). The pipeline additionally provides an automatic classification into object type (galaxy, quasar or star) and a redshift estimate by fitting a model spectrum \citep{2012AJ....144..144B}.

Visual inspection of quasar spectra was an important procedure during the first three generations of SDSS to correct for mis-classifications of object type and inaccurate redshift determinations by the pipeline \citep{2010AJ....139.2360S,2017A&A...597A..79P}. Starting in \mbox{SDSS-IV}, most of the objects are securely classified by the pipeline, with less than 10\% of the spectra requiring visual inspection \citep{2016AJ....151...44D}. The visual-inspection redshifts, when available, are taken as the definitive quasar redshifts, while the remaining quasars have redshifts estimated by the pipeline.

The cross-correlation analysis presented here involves the selection of three quasar samples from DR14Q: tracer quasars (for which we only need the redshifts and positions on the sky), quasars providing Ly$\alpha$ forest absorption in the Ly$\alpha$ region, and quasars providing Ly$\alpha$ forest absorption in the Ly$\beta$ region. The selected sample of tracer quasars contains 266,590 quasars in the range $1.77<z_{\rm q}<3.5$. It includes 13,406 SDSS DR7 quasars \citep{2010AJ....139.2360S} and 18,418 broad absorption line (BAL) quasars, the latter identified as having a CIV balnicity index \citep{1991ApJ...373...23W} \texttt{\detokenize{BI_CIV>}}0 in DR14Q. Quasars with redshifts less than 1.77 are excluded because they are necessarily separated from observable forest pixels (see below) by more than $200~\hMpc$, the maximum distance where the correlation function is measured. The upper limit of $z_{\rm q}=3.5$ is adopted because of the low number of higher-redshift quasars that both limits their usefulness for correlation measurements and make them subject to contamination due to redshift errors of the much more numerous low-redshift quasars \citep{2018arXiv180809955B}. Such contaminations would be expected to add noise (but not signal) to the cross-correlation.

The summary of the Ly$\alpha$ forest data covering the Ly$\alpha$ or Ly$\beta$ region of the quasar spectrum is given in Table~\ref{table::forests}. Both samples exclude SDSS DR7 quasars and BAL quasars. The Ly$\alpha$ sample is derived from a super set consisting of 194,166 quasars in the redshift range $2.05<z_{\rm q}<3.5$, whereas the Ly$\beta$ sample is taken from a super set containing 76,650 quasars with $2.55<z_{\rm q}<3.5$. The lower redshift limits are a consequence of the forests exiting the wavelength coverage of the spectrograph for quasars with $z_{\rm q}<2$ and $z_{\rm q}<2.53$, respectively. Spectra with the same object identification \texttt{\detokenize{THING_ID}} (re-observed quasars) are coadded using inverse-variance weighting. For the selected forest samples, 17\% of the quasars have duplicate spectra (less than 2\% have more than one reobservation) taken with the BOSS spectrograph.

The forest spectra are prepared for analysis by discarding pixels which were flagged as problematic in the flux calibration or sky subtraction by the pipeline. We mask pixels around bright sky lines using the condition $\left| 10^{4}\log_{10}(\lambda/\lambda_{\rm sky})\right|\leq1.5$, where $\lambda_{\rm sky}$ is the wavelength at the pixel center of the sky line where the pipeline sky subtraction is found to be inaccurate. Finally, we double the mask width to remove pixels around the observed CaII H\&K lines arising from absorption by the interstellar medium of the Milky Way.

Forests featuring identified damped Ly$\alpha$ systems (DLAs) are given a special treatment. We use an updated (DR14) version of the DLA catalog of DR9 \citep{2012A&A...547L...1N}. The DLA detection and estimation of the neutral-hydrogen column density $N_{HI}$ was based on correlating observed spectra with synthetic spectra. The effective threshold for DLA detection depends on the signal-to-noise ratio (and therefore on redshift) but is typically $\log_{10}N_{HI}\approx20.3$ for spectra with $S/N>3$ for which the efficiency and purity are $\approx95\%$. For the purposes of the measurement of the correlation function, all pixels in the DLA where the transmission is less  than 20\% are masked and the absorption in the wings is corrected using a Voigt profile following the procedure of \citet{Lee13}. The effect on the correlation function of undetected DLAs or more generally of high-column-density (HCD) systems with $\log_{10}N_{HI}>17.2$ are modeled in the theoretical power spectrum, as described in Sec.~\ref{subsec:otherterm}.

To facilitate the computation of the cross-correlation, we follow the approach in \citet{2017A&A...603A..12B} to combine three adjacent pipeline pixels into wider ``analysis pixels'' defined as the inverse-variance-weighted flux average. Requiring a minimum of 20 analysis pixels in each spectrum discards 2447 (6155) forests for the Ly$\alpha$ (Ly$\beta$) region. Lastly, 3087 (1882) forests failed the continuum-fitting procedure (see section~\ref{section::Measurement_of_the_transmission_field}) for the Ly$\alpha$ (Ly$\beta$) region by having negative continua due to their low spectral signal-to-noise ratios. The final samples include 188,632 forests for the Ly$\alpha$ region and 68,613 forests for the Ly$\beta$ region. Figure~\ref{figure::zhistosamples} shows the redshift distributions for the tracer quasars and the \lya~absorption pixels. Our samples can be compared to those of dMdB17, which included 234,367 quasars (217,780 with $1.8<z_{\rm q}<3.5$) and 168,889 forests (157,845 with $2.0<z_{\rm q}<3.5$) over a wider redshift range.

\begin{table}[tb]
\centering
\caption{Definition of \lya~and \lyb~regions of quasar spectrum in which we measured \lya~forest absorption. The table shows the rest- and observer-frame wavelength ranges defining the regions, the range of quasar redshifts, and the number of forests available in our analysis sample.}
\label{table::forests}
\begin{tabular}{l c c c c}
\hline
\hline
\noalign{\smallskip}
Region & $\lambda_{\rm rf}~[{\rm nm}]$ & $\lambda~[{\rm nm}]$ & $z_{\rm q}$ & $N_{\rm forest}$ \\
\noalign{\smallskip}
\hline
\noalign{\smallskip}
Ly$\alpha$ & $[104,120]$ & $[360,540]$ & $[2.05,3.5]$ & 188,632 \\
Ly$\beta$ & $[97.4,102]$ & $[360,459]$ & $[2.55,3.5]$ & 68,613 \\
\noalign{\smallskip}
\hline
\end{tabular}
\end{table}

%===================================================
\section{The \LyaForest flux-transmission field}
\label{section::Measurement_of_the_transmission_field}
%===================================================

\begin{figure}[tb]
\centering
\includegraphics[width=\columnwidth]{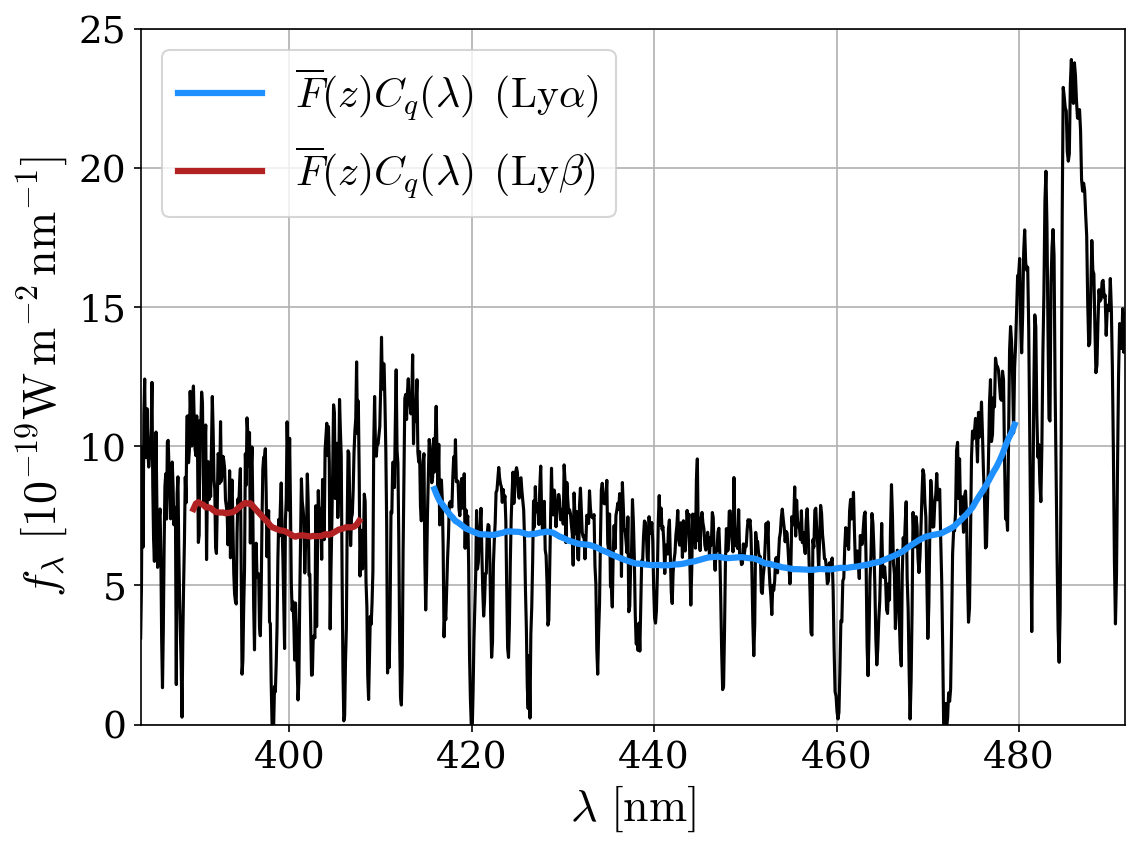}
\caption{Example spectrum of DR14Q quasar identified by $(\mathrm{Plate,MJD,FiberID})=(7305,56991,570)$ at $z_{\rm q}=3.0$. The blue line indicates the best-fit model $\overline{F}(z)C_{\rm q}(\lambda)$ for the Ly$\alpha$ region covering the rest-frame wavelength interval $104.0<\lambda_{\rm rf}<120.0~\mathrm{nm}$. The red line indicates the same for the Ly$\beta$ region over the range $97.4<\lambda_{\rm rf}<102.0~\mathrm{nm}$. The Ly$\alpha$ and Ly$\beta$ emission lines are located at $\lambda_{\alpha}=121.567$~nm and $\lambda_{\beta}=102.572$~nm in the quasar rest-frame. The spectrum has not been rebinned into analysis pixels in this figure.}
\label{figure::spectrum}
\end{figure}

The transmitted flux fraction $F$ in a pixel of the forest region of quasar $q$ is defined as the ratio of the observed flux density $f_{\rm q}$ with the continuum flux $C_{\rm q}$ (the flux density that would be observed in the absence of absorption). We will be studying the transmission relative to the mean value at the observed wavelength $\overline{F}(\lambda)$, and refer to this quantity as the ``delta-field'':
\begin{equation}
\delta_{\rm q}(\lambda) = \frac{f_{\rm q}(\lambda)}{C_{\rm q}(\lambda)\overline{F}(\lambda)}-1\ .
\label{eqn:deltafield}
\end{equation}

We employ a similar method to the one established by previous \LyaForest BAO analyses \citep{2013A&A...552A..96B,2015A&A...574A..59D} in which the delta-field is derived by estimating the product $C_{\rm q}(\lambda)\overline{F}(\lambda)$ for each quasar. Each spectrum is modeled assuming a uniform forest spectral template which is multiplied by a quasar-dependent linear function, setting the overall amplitude and slope, to account for the diversity of quasar luminosity and spectral shape:
\begin{equation}
C_{\rm q}(\lambda)\overline{F}(\lambda) = \overline{f}(\lambda_{\rm rf})(a_{\rm q}+b_{\rm q}\log_{10}(\lambda))\ ,
\end{equation}
where $a_{\rm q}$ and $b_{\rm q}$ are free parameters fit to the observed flux of the quasar. The forest spectral template $\overline{f}(\lambda_{\rm rf})$ is derived from the data as a weighted mean normalized flux, obtained by stacking the spectra in the quasar rest-frame. The continuum fitting procedure is handled separately for the Ly$\alpha$ and Ly$\beta$ regions.

The total variance of the delta-field is modeled as
\begin{equation}
\sigma^{2}(\lambda) = \eta(\lambda)\sigma_{\rm noise}^{2}(\lambda)+\sigma_{\rm LSS}^{2}(\lambda)+\epsilon(\lambda)/\sigma_{\rm noise}^{2}(\lambda)\ ,
\label{eq:variance}
\end{equation}
where the noise variance $\sigma_{\rm noise}^{2}=\sigma_{\rm pipe}^{2}/(C_{\rm q}\overline{F})^{2}$. The first term represents the pipeline estimate of the flux variance, corrected by a function $\eta(\lambda)$ that accounts for possible misestimation. The second term gives the contribution due to the large-scale structure (LSS) and acts as a lower limit on the variance at high signal-to-noise ratio. Lastly, the third term absorbs additional variance from quasar diversity apparent at high signal-to-noise ratio. In bins of $\sigma_{\rm noise}^{2}$ and observed wavelength, we measure the variance of the delta-field and fit for the values of $\eta$, $\sigma_{\rm LSS}^{2}$ and $\epsilon$ as a function of observed wavelength. These three functions are different for the Ly$\alpha$ and Ly$\beta$ regions. The procedure of stacking the spectra, fitting the continua and measuring the variance of $\delta$ is iterated, until the three functions converge. We find that five iterations is sufficient. Figure~\ref{figure::spectrum} presents an example spectrum and the best-fit model $C_{\rm q}(\lambda)\overline{F}(\lambda)$ for the Ly$\alpha$ and Ly$\beta$ regions.

As detailed in \citet{2017A&A...603A..12B}, the delta-field can be redefined in two steps to make exact the biases introduced by the continuum fitting procedure. In the first step, we define
\begin{equation}
\hat\delta_{\rm q}(\lambda) = \delta_{\rm q}(\lambda) - \overline{\delta_{\rm q}} - (\Lambda-\overline{\Lambda_{\rm q}})\frac{\overline{(\Lambda-\overline{\Lambda_{\rm q}})\delta_{\rm q}}}{\overline{(\Lambda-\overline{\Lambda_{\rm q}})^{2}}}
\quad , \quad
\Lambda\equiv\log_{10}(\lambda)\ ,
\label{eq::deltaredefine}
\end{equation}
where the over-bars refer to weighted averages over individual forests. Next, we transform the $\hat\delta_{\rm q}(\lambda)$ by subtracting the weighted average at each observed wavelength:
\begin{equation}
  \hat\delta_{\rm q}(\lambda)\rightarrow\hat\delta_{\rm q}(\lambda)-\overline{\delta(\lambda)}\ .
\label{eq::deltaredefine2}
\end{equation}

%===================================================
\section{The \LyaForest - quasar cross-correlation}
\label{section::The_Lya_forest_quasar_cross_correlation}
%===================================================

\begin{table}[tb]
\centering
\caption{Parameters of flat $\Lambda$CDM fiducial cosmological model \citep{2016A&A...594A..13P}. The sound horizon at the drag epoch, $r_d$, is calculated using CAMB \citep{2000ApJ...538..473L}. The Hubble distance $D_{H}$ and the comoving angular diameter distance $D_{M}$ relative to $r_{d}$ are given at the effective redshift of the measurement $z_{\rm eff}$.}
\label{table::cosmology}
\begin{tabular}{l c}
\hline
\hline
\noalign{\smallskip}
Parameter & Value \\
\noalign{\smallskip}
\hline
\noalign{\smallskip}
$\Omega_{c} h^2$ & 0.1197 \\
$\Omega_{b} h^2$ & 0.02222 \\
$\Omega_{\nu} h^2$ & 0.0006 \\
$h$ & 0.6731 \\
$\Omega_{m}=\Omega_{c}+\Omega_{b}+\Omega_{\nu}$ & 0.3146 \\
$n_{s}$ & 0.9655 \\
$\sigma_{8}$ & 0.8298 \\
$N_{\nu}$ & 3	 \\
\noalign{\smallskip}
\hline
\noalign{\smallskip}
$r_{d}~[\hMpc]$ & 99.17 \\
$r_{d}~[\mathrm{Mpc}]$ & 147.33 \\
$z_{\rm eff}$ & 2.35 \\
$D_{H}(z_{\rm eff})/r_{d}$ & 8.55 \\
$D_{M}(z_{\rm eff})/r_{d}$ & 39.35 \\
\noalign{\smallskip}
\hline
\end{tabular}
\end{table}

\begin{figure}[tb]
\centering
\includegraphics[width=\columnwidth]{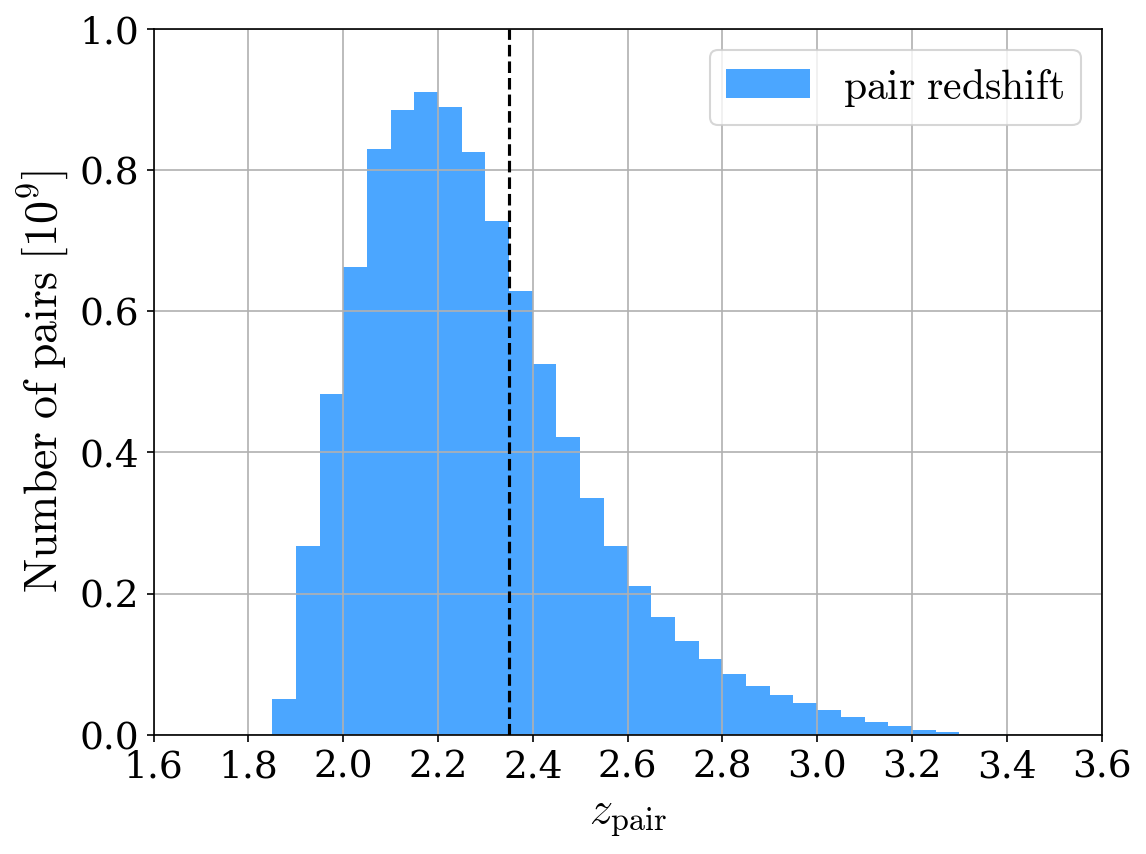}
\caption{Redshift distribution of $9.7\times10^{9}$ correlation pairs. The dashed vertical black line indicates the effective redshift of the BAO measurement, $z_{\rm eff}=2.35$, calculated as the weighted mean of the pair redshifts for separations in the range $80<r<120~\hMpc$.}
\label{figure::zhistopairs}
\end{figure}

The three-dimensional positions of the quasars and the Ly$\alpha$ forest delta-field are determined by their redshifts and angular positions on the sky. We transform the observed angular and redshift separations ($\Delta\theta,\Delta z$) of the quasar-\lya~absorption pixel pairs into Cartesian coordinates ($r_{\perp},r_{\parallel}$) assuming a spatially flat fiducial cosmology. The comoving separations along the line of sight $r_{\parallel}$ (parallel direction) and transverse to the line of sight $r_{\perp}$ (perpendicular direction) are calculated as
\begin{equation}
r_{\parallel} = (D_{\alpha}-D_{\rm q})\cos\left(\frac{\Delta\theta}{2} \right)
\end{equation}
\begin{equation}
r_{\perp} = (D_{\alpha}+D_{\rm q})\sin\left(\frac{\Delta\theta}{2} \right)\ ,
\end{equation}
where $D_{\alpha}\equiv D_{c}(z_{\alpha})$ and $D_{\rm q}\equiv D_{c}(z_{\rm q})$ are the comoving distances to the \lya~absorption pixel and the quasar, respectively. Line of sight separations $r_{\parallel}>0$ ($<0$) thus correspond to background (foreground) absorption with respect to the tracer quasar position. In this paper, we will also refer to the coordinates $(r,\mu)$, where $r^{2} = r_{\parallel}^{2}+r_{\perp}^{2}$ and $\mu = r_{\parallel}/r$, the cosine of the angle of the vector $\mathbf{r}$ from the line of sight. The pair redshift is defined as $z_{\rm pair}=(z_{\alpha}+z_{\rm q})/2$. A histogram of the pair redshifts is displayed in Fig.~\ref{figure::zhistopairs}. We do not include pairs involving a quasar and pixels from its own forest in the cross-correlation analysis, because the correlation of such pairs vanishes due to the continuum fit and delta-field redefinition (eqn~\ref{eq::deltaredefine}).

The fiducial cosmology used in the analysis is a flat $\Lambda$CDM model with parameter values taken from the Planck (2016) result for the TT+lowP combination \citep{2016A&A...594A..13P} described in Table~\ref{table::cosmology}. It is the same fiducial cosmology employed by dMdB17.

%--------------------------------------
\subsection{Cross-correlation}
%--------------------------------------
\label{subsec:cross}

We estimate the cross-correlation at a separation bin $A$, $\xi_{A}$, as the weighted mean of the delta-field in pairs of pixel $i$ and quasar $k$ at a separation within the bin $A$ \citep{2012JCAP...11..059F}:
\begin{equation}
\hat{\xi}_{A}=\frac{\sum\limits_{(i,k)\in A}w_{i}\delta_{i}}{\sum\limits_{(i,k)\in A}w_{i}}\ .
\end{equation}
The weights $w_{i}$ are defined as the inverse of the total pixel variance (see equation~\ref{eq:variance}), multiplied by redshift evolution factors for the forest and quasar, so as to approximately minimize the relative error on $\hat{\xi}_{A}$ \citep{2013A&A...552A..96B}:
\begin{equation}
w_{i} = \sigma_{i}^{-2}\left(\frac{1+z_{i}}{3.25}\right)^{\gamma_{\alpha}-1}\left(\frac{1+z_{k}}{3.25}\right)^{\gamma_{\rm q}-1}\ ,
\label{eq:weights}
\end{equation}
where $\gamma_{\alpha}=2.9$ \citep{2006ApJS..163...80M} and $\gamma_{\rm q}=1.44$ \citep{2019ApJ...878...47D}. The validity of the correlation estimator, as well as the accuracy of the distortion matrix (section~\ref{subsec:distortion}) and covariance matrix estimation (section~\ref{subsec:covariance}) were tested and confirmed on simulated data in dMdB17.

Our separation grid consists of 100 bins of $4~\hMpc$ for separations $r_{\parallel}\in[-200,200]~\hMpc$ in the parallel direction and 50 bins of $4~\hMpc$ for separations $r_{\perp}\in[0,200]~\hMpc$ in the perpendicular direction; the total number of bins is $N_{\rm bin}=5000$. Each bin is defined by the weighted mean ($r_{\perp},r_{\parallel}$) of the quasar-pixel pairs of that bin, and its redshift by the weighted mean pair redshift. The mean redshifts range from $z=2.29$ to $z=2.40$. The effective redshift of the cross-correlation measurement is defined to be the inverse-variance-weighted mean of the redshifts of the bins with separations in the range $80<r<120~\hMpc$ around the BAO scale. Its value is $z_{\rm eff}=2.35$.

Because the Ly$\beta$ transition is sufficiently separated in wavelength from the Ly$\alpha$ transition, corresponding to large physical separations $>441~\hMpc$ for the wavelength range of the analysis, we neglect the contamination from Ly$\beta$ absorption interpreted as Ly$\alpha$ absorption. The total number of pairs of the cross-correlation measurement is $9.7\times10^{9}$. The Ly$\alpha$ absorption in the Ly$\beta$ region contributes $1.2\times10^{9}$ pairs (13\%) and reduces the mean variance of the correlation function by 9\% compared to the Ly$\alpha$ region-only measurement. Our cross-correlation measurement has 39\% lower mean variance than the measurement of dMdB17.

%--------------------------------------
\subsection{Distortion matrix}
%--------------------------------------
\label{subsec:distortion}

The procedure used to estimate the delta-field (section~\ref{section::Measurement_of_the_transmission_field}) suppresses fluctuations of characteristic scales corresponding to the forest length, since the estimate of the product $C\overline{F}$ (equation~\ref{eqn:deltafield}) would typically erase such a fluctuation. The result is a suppression of the power spectrum in the radial direction
on large scales (low $k_\parallel$). As illustrated in Figure 11 of dMdB17, this induces a significant but smooth distortion of the correlation function on all relevant scales while leaving the BAO peak visually intact. As first noted in \citet{2011JCAP...09..001S} and further investigated in \citet{2015JCAP...11..034B}, the distortion effect can be modeled in Fourier space as a multiplicative function of the radial component $k_{\parallel}$ on the Ly$\alpha$ forest transmission power spectrum.

Here, we use the method introduced by \citet{2017A&A...603A..12B} for the Ly$\alpha$ auto-correlation and adapted to the cross-correlation by dMdB17 which allows one to encode the effect of this distortion on the correlation function in a distortion matrix. This approach, extensively validated in these publications using simulated data, uses the fact that that equations (\ref{eq::deltaredefine}) and
(\ref{eq::deltaredefine2}) are linear in $\delta$. This fact allows one
to describe the measured correlation function for a separation bin A as a linear combination of the true correlation function for bins $A^\prime$:

\begin{equation}
\hat\xi_{A} = \sum\limits_{A^\prime}D_{AA^\prime}\xi_{ A^\prime}\ .
\end{equation}
The distortion matrix $D_{AA^\prime}$ depends only on the geometry of the survey, the lengths of the forests and the pixel weights,
\begin{equation}
D_{AA^\prime} = \frac{\sum\limits_{(i,k)\in A}w_{i}\sum\limits_{(j,k)\in A^{\prime}}P_{ij}}{\sum\limits_{(i,k)\in A}w_{i}}\ ,
\label{equation::distortionmatrix}
\end{equation}
where the projection matrix
\begin{equation}
P_{ij} = \delta_{ij}^{K}-\frac{w_{j}}{\sum\limits_{l}w_{l}}-\frac{w_{j}(\Lambda_{i}-\overline{\Lambda})(\Lambda_{j}-\overline{\Lambda})}{\sum\limits_{l}w_{l}(\Lambda_{l}-\overline{\Lambda})^{2}}\quad , \quad
\Lambda\equiv\log_{10}(\lambda)\ ,
\end{equation}
and $\delta^K$ is the Kronecker delta. The indices $i$ and $j$ in equation~ (\ref{equation::distortionmatrix}) refer to pixels from the same forest, $k$ refers to a quasar, and the sums run over all pixel-quasar pairs that contribute to the separation bins $A$ and $A'$. The diagonal elements dominate the distortion matrix and are close to unity, $D_{AA}\approx0.97$, whereas the off-diagonal elements are small, $\left| D_{AA^\prime}\right| \lesssim0.03$. We use the distortion matrix when performing fits of the measured cross-correlation function (see equation~\ref{equation::xi_complete}).

%--------------------------------------
\subsection{Covariance matrix}
%--------------------------------------
\label{subsec:covariance}

\begin{figure*}[tb]
\centering
\includegraphics[width=\columnwidth]{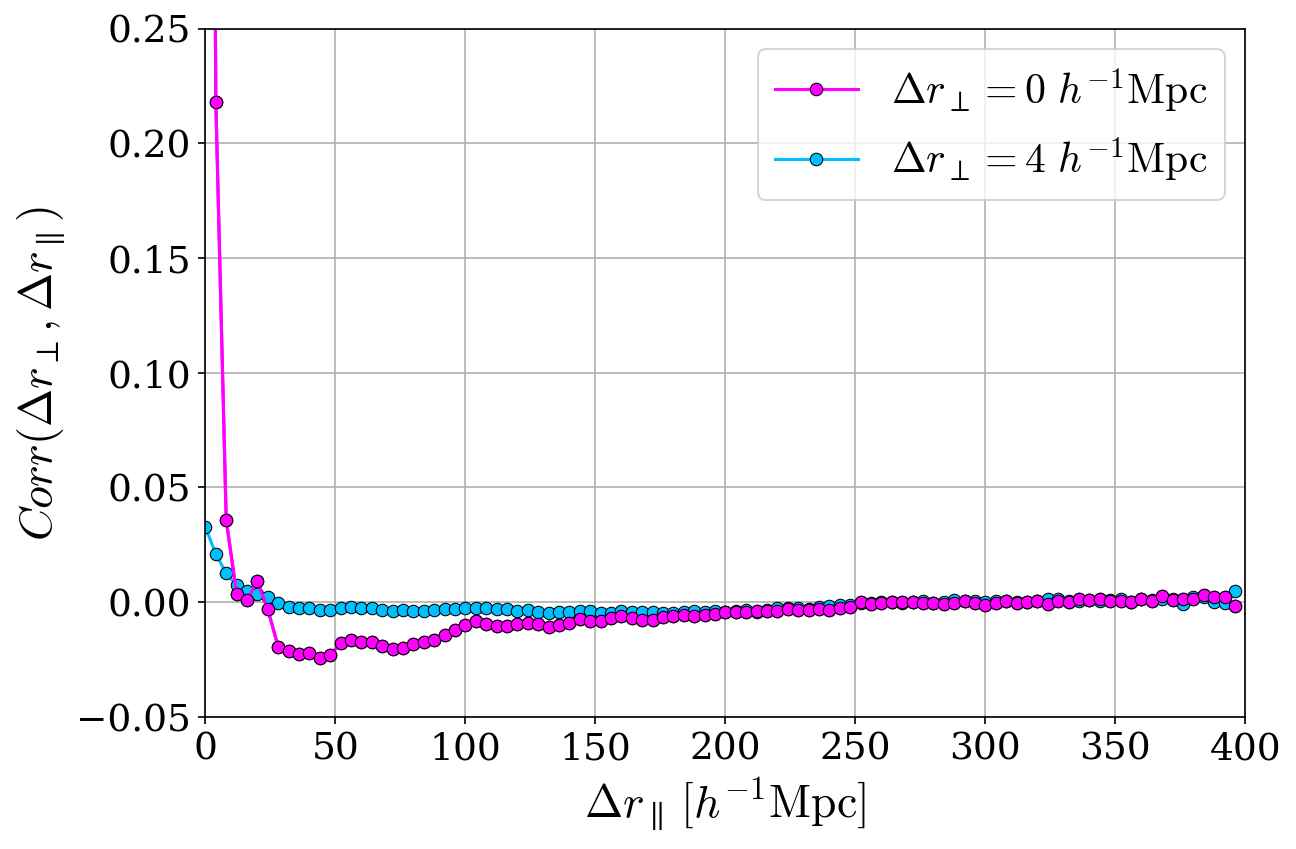}
\includegraphics[width=\columnwidth]{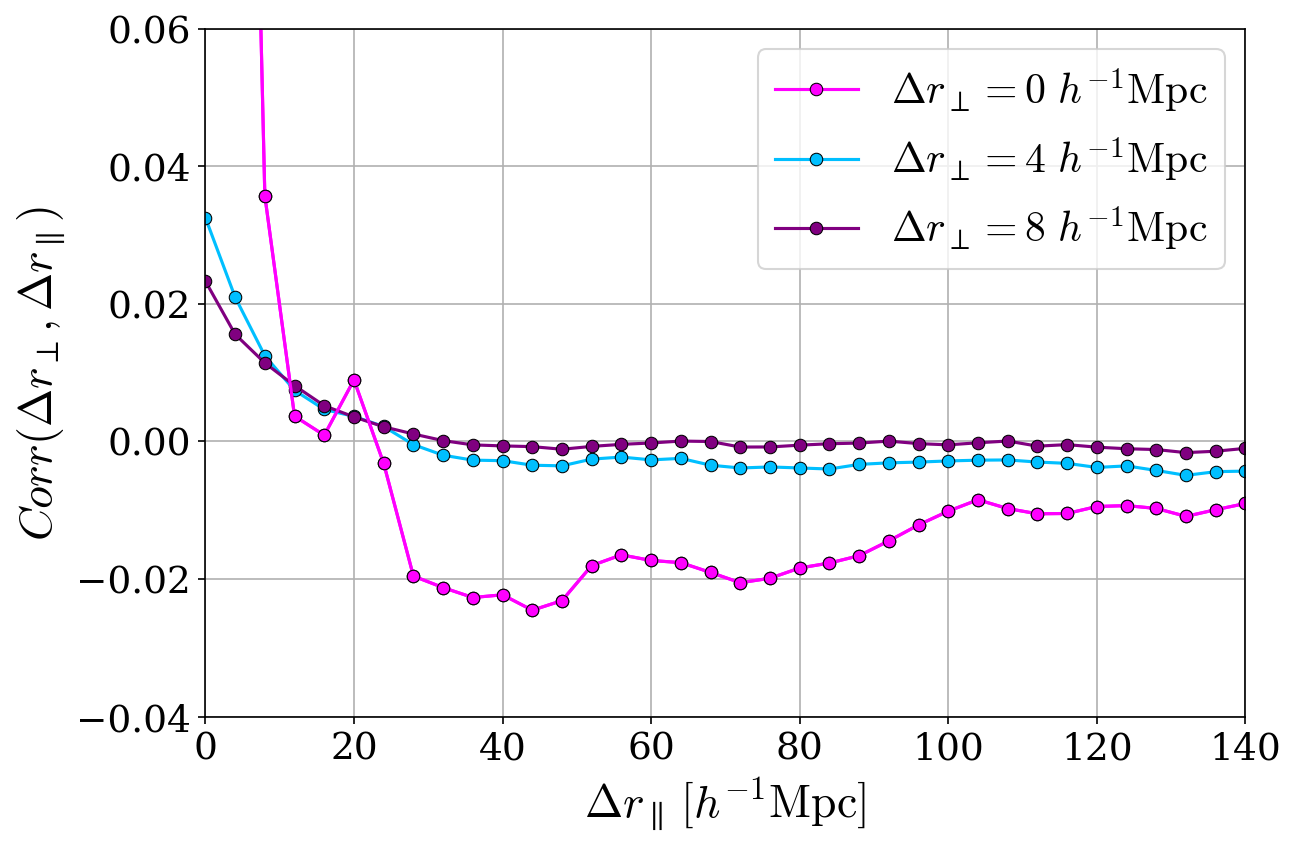}
\caption{Smoothed correlation matrix from sub-sampling as a function of $\Delta r_{\parallel}=|r_{\parallel,A}-r_{\parallel,B}|$. The curves are for constant $\Delta r_{\perp}=|r_{\perp,A}-r_{\perp,B}|$ for the three lowest values $\Delta r_{\perp}=\left[0,4,8\right]~\hMpc$. The right panel shows an expansion of the region $\Delta r_{\parallel}<140~\hMpc$.}
\label{figure::cov}
\end{figure*}

We estimate the covariance matrix of the cross-correlation from the data by using the subsampling technique introduced by \citet{2013A&A...552A..96B} and adapted to the cross-correlation by dMdB17. We divide the DR14 footprint of Figure~\ref{figure::sky} into subsamples and measure the covariance from the variability across the subsamples. Such estimates of the covariance matrix are unbiased, but the noise due to the finite number of subsamples leads to biases in the inverse of the covariance \citep{2014IAUS..306...99J}. As was done in dMdB17, we smooth the noise by assuming, to good approximation, that the covariance between separation bins $A$ and $B$ depends only on the absolute difference $(\Delta r_{\parallel},\Delta r_{\perp})=(|r_\parallel^A-r_\parallel^B|,|r_\perp^A-r_\perp^B|)$.

We define the subsamples through a HEALPix \citep{2005ApJ...622..759G} pixelization of the sky. A quasar-absorption pixel pair is assigned to a subsample $s$ if the forest that contains the absorption belongs to that HEALPix pixel. We use HEALPix parameter \texttt{nside}=32, resulting in 3262 subsamples. Using fewer but larger HEALPix pixels (\texttt{nside}=16, 876 subsamples) has no significant impact on the covariance matrix or the BAO peak position measurement.

The (noisy) covariance matrix is calculated as
\begin{equation}
C_{AB} = \frac{1}{W_{A}W_{B}}\sum\limits_{s}W_{A}^{s}W_{B}^{s}\left[\xi_{A}^{s}\xi_{B}^{s}-\xi_{A}\xi_{B}\right]\ ,
\end{equation}
where the sum runs over all subsamples and $W_{A}$ is the sum of the pair weights $w$ belonging to bin $A$,
\begin{equation}
W_{A}=\sum\limits_{i\in A}w_{i}\ .
\end{equation}
From the covariance, we calculate the correlation matrix:
\begin{equation}
Corr_{AB}=\frac{C_{AB}}{\sqrt{C_{AA}C_{BB}}}\ .
\end{equation}
The smoothing procedure is applied to this correlation matrix by averaging as a function of $(\Delta r_{\parallel},\Delta r_{\perp})$. The final covariance used in the fits is obtained by multiplying the smoothed correlation matrix by the diagonal elements of the original covariance matrix. Figure~\ref{figure::cov} displays the smoothed correlation matrix as a function of $\Delta r_{\parallel}$ for the three lowest values of $\Delta r_{\perp}$.

%===================================================
\section{Model of the cross-correlation}
\label{section::model}
%===================================================

\begin{table*}[tb]
\centering
\caption{List of parameters of cross-correlation model. The 14 parameters of the standard fit are given in the first section of the table. The second section lists parameters that are fixed in the standard fit.
} 
\label{table::parameters}
\begin{tabular}{l l}
\hline
\hline
\noalign{\smallskip}
Parameter & Description\\
\noalign{\smallskip}
\hline
\noalign{\smallskip}
$\alpha_{\parallel}, \alpha_{\perp}$ & BAO peak-position parameters\\
$\beta_{\alpha}$ & Redshift-space distortion parameter for Ly$\alpha$ absorption\\
$b_{\eta\alpha}$ & Velocity gradient bias for Ly$\alpha$ absorption\\
$\sigma_{v}~[\hMpc]$ & Smoothing parameter for quasar nonlinear velocities and redshift precision\\
$\Delta r_{\parallel}~[\hMpc]$ & Coordinate shift due to the quasar redshift systematic error\\
$\xi^{\rm TP}_{0}$ & Amplitude parameter of quasar radiation\\
$b_{\rm HCD}^{\rm Ly\alpha-quasar}$ & Bias parameter of HCD systems\\
$\beta_{\rm HCD}$ & Redshift-space distortion parameter of HCD systems\\
$A_{\rm rel1}$ & Dipole amplitude of relativistic correction\\
$b_{\rm m}$ & Transmission bias parameters of four metal species\\
\noalign{\smallskip}
\hline
\noalign{\smallskip}
$b_{\rm q}=3.77$ & Bias parameter for quasars\\
$\beta_{\rm q}=0.257$ & Redshift-space distortion parameter for quasars\\
$f=0.969$ & Growth rate of structure\\
$\Sigma_{\perp}=3.26~\hMpc$ & Transverse nonlinear broadening of the BAO peak\\
$1+f=1.969$ & Ratio of radial to transverse nonlinear broadening\\
$L_{\rm HCD}=10~\hMpc$ & Smoothing scale of HCD systems\\
$\lambda_{\rm UV}=300~\hMpc$ & Mean free path of UV photons\\
$A_{\rm rel3}=0$ & Octupole amplitude of relativistic correction\\
$\beta_{\rm m}=0.5$ & Redshift-space distortion parameters of four metal species\\
$R_{\parallel}=4~\hMpc$ & Radial binning smoothing parameter\\
$R_{\perp}=4~\hMpc$ & Transverse binning smoothing parameter\\
$A_{\rm peak}=1$ & BAO peak amplitude\\
$\gamma_{\alpha}=2.9$ & Ly$\alpha$ transmission bias evolution exponent\\
$\gamma_{\rm m}=1$ & Metal transmission bias evolution exponent\\
\noalign{\smallskip}
\hline
\end{tabular}
\end{table*}

We fit the measured cross-correlation function, $\hat{\xi}_A$, in the $(\rperp,\rpar)$ bin $A$, to a cosmological correlation
function $\xi_{A^\prime}^{\rm cosmo}$:
\begin{equation}
  \hat{\xi}_A = \sum\limits_{A^\prime} D_{AA^\prime}\left[ \xi_{A^\prime}^{\rm cosmo} + \xi_{A^\prime}^{\rm bb} \right]\ ,
  \label{equation::xi_complete}
\end{equation}
where $D_{AA^\prime}$ is the distortion matrix (equation~\ref{equation::distortionmatrix}). The broadband term, $\xi_{A}^{\rm bb}$, is an optional function used to test for imperfections in the model and for systematic errors. The set of parameters for the model is summarized in Table \ref{table::parameters}. The model is calculated at the weighted mean $(r_{\perp},r_{\parallel})$ and redshift of each bin of the correlation function. Because of the relatively narrow redshift distribution of the bins $(\Delta z=0.11)$, most model parameters can be assumed as redshift independent to good accuracy.

The cosmological cross-correlation function is the sum of several
contributions
\begin{equation}
  \xi^{\rm cosmo} = \xi^{\rm q\alpha}+\sum\limits_{\rm m} \xi^{\rm qm}+\xi^{\rm qHCD}+\xi^{\rm TP}+\xi^{\rm rel}+\xi^{\rm asy}\ .
  \label{equation::theoretical_cosmo}
\end{equation}
The first term represents the standard correlation between quasars, $q$, and \lya~absorption in the IGM. It is the most important part of the correlation function and, used by itself, would lead to an accurate determination of the BAO peak position (see results in section~\ref{section::fits}).

The remaining terms in equation~(\ref{equation::theoretical_cosmo}) represent
subdominant effects but contribute toward improving the fit of the correlation function outside the BAO peak. The second term is the sum over correlations from metal absorbers in the IGM. The third term represents \lya~absorption by high column density systems (HCDs). The fourth term is the correlation from the effect of a quasar's radiation on a neighboring forest (``transverse proximity effect''). The fifth term is a relativistic correction leading to odd-$\ell$ multipoles in the correlation function, and the final term includes other sources of odd-$\ell$ multipoles \citep{2014PhRvD..89h3535B}. These terms will be described in detail below. 

%--------------------------------------
\subsection{Quasar-Ly$\alpha$ correlation term}
%--------------------------------------
\label{subsec:qlyaterm}

The quasar-\lya~cross-correlation, $\xi^{\rm q\alpha}$, is the dominant contribution to the cosmological cross-correlation. It is assumed to be a biased version of the total matter auto-correlation of the appropriate flat \lcdm~model, separated into a smooth component and a peak component to free the position of the BAO peak:
\begin{equation}
  \xi^{\rm q\alpha}(r_{\perp},r_{\parallel},\alpha_{\perp},\alpha_{\parallel}) = \xi_{\rm sm}( r_{\perp},r_{\parallel}) + A_{\rm peak}\xi_{\rm peak} (\alpha_{\perp}r_{\perp},\alpha_{\parallel}r_{\parallel})\ ,
  \label{equation::xismoothpeak}
\end{equation}
where $A_{\rm peak}$ is the BAO peak amplitude. The anisotropic shift of the observed BAO peak position relative to the peak position of the fiducial cosmological model from Table~\ref{table::cosmology} is described by the line-of-sight and transverse scale parameters
\begin{equation}
\alpha_{\parallel} = \frac {\left[D_{H}(z_{\rm eff})/r_{d}\right]}
{\left[D_{H}(z_{\rm eff})/r_{d}\right]_{\rm fid}}
\hspace*{3mm}{\rm and}\hspace*{5mm}
\alpha_{\perp} = \frac {\left[D_{M}(z_{\rm eff})/r_{d}\right]}
{\left[D_{M}(z_{\rm eff})/r_{d}\right]_{\rm fid}}\ .
\end{equation}

The nominal correlation function,
$\xi^{\rm q\alpha}(\rperp,\rpar,\aperp=\apar=1)$, is the Fourier transform of the quasar-\lya~cross-power spectrum:
\begin{equation}
P^{\rm q\alpha}(\vec{k},z) = P_{\rm QL}(\vec{k},z)d_{\rm q}(\mu_{k},z)d_{\alpha}(\mu_{k},z)\sqrt{V_{\rm NL}(k_{\parallel})}G(\vec{k})\ ,
\label{equation::pk_1}
\end{equation}
where $\vec{k} = (k_{\parallel},k_{\perp})$ is the wavenumber of modulus $k$ with components $k_{\parallel}$ along the line of sight and $k_{\perp}$ across, and $\mu_{k} = k_{\parallel}/k$ is the cosine of the angle of the wavenumber from the line of sight.
As described in detail below, $P_{\rm QL}$ is the (quasi) linear matter spectrum, $d_q$ and $d_{\rm Ly\alpha}$ are 
the standard linear-theory factors describing the tracer bias and redshift-space distortion \citep{1987MNRAS.227....1K}, $V_{\rm NL}$ describes further nonlinear corrections not included in $P_{\rm QL}$, and $G(\vec{k})$ gives the effects of $(\rperp,\rpar)$ binning on the measurement.

The first term in (\ref{equation::pk_1}) provides for the aforementioned decoupling of the peak component (Eq. \ref{equation::xismoothpeak}):
\begin{equation}
P_{\rm QL}(\vec{k},z) = P_{\rm sm}(k,z) + \exp\left[-(k_{\parallel}^2\Sigma_{\parallel}^2+k_{\perp}^2\Sigma_{\perp}^2)/2\right]P_{\rm peak}(k,z)\ ,
\end{equation}
where the smooth component, $P_{\rm sm}$, is derived from the linear power spectrum, $P_{\rm L}(k,z)$, via the side-band technique \mbox{\citep{2013JCAP...03..024K}} and $P_{\rm peak}=P_{\rm L}-P_{\rm sm}$. The redshift-dependent linear power spectrum is obtained from CAMB \citep{2000ApJ...538..473L} with the fiducial cosmology.

The correction for nonlinear broadening of the BAO peak is parameterized by
$\vec{\Sigma}=(\Sigma_{\parallel},\Sigma_{\perp})$,
with $\Sigma_\perp=3.26~\hMpc$ and
\begin{equation}
\frac{\Sigma_{\parallel}}{\Sigma_{\perp}}=1+f\ ,
\end{equation}
where $f=d(\ln g)/d(\ln a)\approx\Omega_{\rm m}^{0.55}(z)$ is the linear growth rate of structure.

The second term in (\ref{equation::pk_1}) describes the quasar
bias and redshift-space distortion
\begin{equation}
d_{\rm q}(\mu_{k},z) = b_{\rm q}(z)\left(1 + \beta_{\rm q}\mu_{k}^{2}\right)\ .
\label{equation::dq}
\end{equation}
Because the fit of the cross-correlation is only sensitive to the product of the quasar and Ly$\alpha$ biases, we set $b_{\rm q}\equiv b_{\rm q}(z_{\rm eff}) = 3.77$ and assume a redshift dependence of the quasar bias given by \citep{2005MNRAS.356..415C}
\begin{equation}
b_{\rm q}(z) = 0.53 + 0.289(1+z)^{2}\ .
\end{equation}
The quasar redshift-space distortion, assumed to be redshift independent, is
\begin{equation}
\beta_{\rm q} = \frac{f}{b_{\rm q}}\ .
\end{equation}
Setting $f=0.969$ for our fiducial cosmology yields $\beta_{\rm q}=0.257$.

The third term in (\ref{equation::pk_1}) is the \LyaForest factor,
\begin{equation}
d_{\alpha}(\mu_{k},z) = b_{\alpha}(z)\left(1 + \beta_{\alpha}\mu_{k}^{2}\right)\ .
\label{equation::dalpha}
\end{equation}
We assume that the transmission bias evolves with redshift as
\begin{equation}
b_{\alpha}(z) = b_{\alpha}(z_{\rm eff})\left( \frac{1+z}{1+z_{\rm eff}} \right)^{\gamma_{\alpha}}\ ,
\end{equation}
with $\gamma_{\alpha}=2.9$ \citep{2006ApJS..163...80M}, while $\beta_{\alpha}$ is assumed to be redshift independent. We choose to fit for $\beta_{\alpha}$ and the velocity gradient bias of the Ly$\alpha$ forest:
\begin{equation}
b_{\eta\alpha}=b_{\alpha}\beta_{\alpha}/f\ .
\end{equation}

Beyond our standard treatment of the Ly$\alpha$ transmission bias, we also consider the effect of fluctuations of ionizing UV radiation which lead to a scale-dependence of $b_\alpha$ \citep{2014PhRvD..89h3010P,2014MNRAS.442..187G}:
\begin{equation}
b_{\alpha}(k) = b_{\alpha}+b_{\Gamma}\frac{W(k\lambda_{\rm UV})}{1+b_{a}^{\prime}W(k\lambda_{\rm UV})}\ ,
\label{equation::uv}
\end{equation}
where $W(x)=\arctan(x)/x$ \citep[following the parameterization of][]{2014MNRAS.442..187G}. Our standard fit does not include the effect of UV fluctuations due to its minor contribution to the fit quality. A fit that includes the UV modeling is presented in Table~\ref{table::nonstandard} for which we fix the UV photon mean free path $\lambda_{\rm UV}=300~\hMpc$ \citep{2013ApJ...769..146R} and $b_a^\prime=-2/3$ \citep{2014MNRAS.442..187G}, and fit for $b_\Gamma$, as was done in dMdB17.

The effect of quasar nonlinear velocities and statistical redshift errors on the power spectrum is modeled as a Lorentz damping \citep{2009MNRAS.393..297P},
\begin{equation}
V_{\rm NL}(k_{\parallel}) = \frac{1}{1+(k_{\parallel}\sigma_{v})^2}\ ,
\end{equation}
where $\sigma_v$ is a free parameter.

The last term in (\ref{equation::pk_1}), $G(\vec{k})$, accounts for smoothing due to the binning of the measurement of the correlation function \citep{2017A&A...603A..12B}. We use
\begin{equation}
G(\vec{k}) = \sinc{\left(\frac{k_{\parallel}R_{\parallel}}{2}\right)}\sinc{\left(\frac{k_{\perp}R_{\perp}}{2}\right)}\ ,
\end{equation}
where $R_{\parallel}$ and $R_{\perp}$ are the  scales of the smoothing. In the transverse direction, this form is not exact, but we have verified that it generates a sufficiently accurate correlation function. We fix both to the bin width, $R_{\parallel}=R_{\perp}=4~\hMpc$.

Systematic errors in the quasar redshift estimates lead to a shift of the cross-correlation along the line of sight which is accounted for in the fit using the free parameter
\begin{equation}
\Delta r_{\parallel} =  r_{\parallel,{\rm true}}-r_{\parallel,{\rm measured}} = \frac{(1+z)\Delta v_{\parallel}}{H(z)}\ .
\end{equation}

%--------------------------------------
\subsection{Quasar-metal correlation terms}
%--------------------------------------
\label{subsec:qmetterm}

\begin{table}[tb]
\centering
\caption{Most important metal absorptions of intergalactic medium that imprint correlations observed in Ly$\alpha$-quasar cross-correlation for $r_{\parallel}\in [-200,200]~\hMpc$. The second column lists the rest-frame wavelength of the metal line and the third column its ratio with $\lambda_{\alpha}$ (using the shorter of the two wavelengths in the denominator). The last column gives the apparent radial distance difference between the Ly$\alpha$ and metal absorption, $r_{\parallel}=D_{c}(z_{\alpha})-D_{c}(z_{\rm m})$, for observed wavelength $\lambda=407.2$~nm (corresponding to Ly$\alpha$ absorption at $z_{\rm eff}=2.35$).}
\label{table::metals}
\begin{tabular}{l c c c}
\hline
\hline
\noalign{\smallskip}
Metal line & $\lambda_{\rm m}~[{\rm nm}]$ & $\lambda_{1}/\lambda_{2}$ & $r_{\parallel}^{\rm \alpha m}~[\hMpc]$ \\
\noalign{\smallskip}
\hline
\noalign{\smallskip}
SiII(119.0) & 119.04 & 1.021 & -59 \\
SiII(119.3) & 119.33 & 1.019 & -53 \\
SiIII(120.7) & 120.65 & 1.008 & -21 \\
SiII(126.0) & 126.04 & 1.037 & +103 \\
\noalign{\smallskip}
\hline
\end{tabular}
\end{table}

Absorption by metals in the intergalactic medium (e.g., \citealt{2014MNRAS.441.1718P}) with similar rest-frame wavelengths to Ly$\alpha$ yields a sub-dominant contribution to the measured cross-correlation. Assuming that these contaminant absorptions have redshifts corresponding to Ly$\alpha$ absorption results in an apparent shift of the quasar-metal cross-correlations along the line of sight in the observed cross-correlation. Following \citet{2018JCAP...05..029B}, metal correlations are modeled as
\begin{equation}
\xi^{{\rm qm}}_{A} = \sum_{B} M_{AB} \xi^{{\rm qm}}_{B}\ ,
\end{equation}
where
\begin{equation}
M_{AB} \equiv \frac{1}{W_{A}}\sum_{(i,k)\in A, (i,k)\in B}w_{i}
\end{equation}
is a ``metal distortion matrix'' that allows us to calculate the shifted quasar-metal cross-correlation function for a given non-shifted quasar-metal cross-correlation function. The condition $(i,k)\in A$ refers to pixel distances calculated using $z_{\alpha}$, but $(i,k)\in B$ refers to pixel distances calculated using $z_{\rm{m}}$. For each metal absorption line, the (non-shifted) quasar-metal correlation is modeled using (\ref{equation::pk_1}) with $d_{\alpha}$ replaced by
\begin{equation}
d_{\rm m}(\mu_{k},z) = b_{\rm m}(z)\left(1 + \beta_{\rm m}\mu_{k}^{2}\right)\ .
\end{equation}
The metal absorption lines included in the fit are listed in Table~\ref{table::metals}. Because the redshift-space distortion parameter of each metal is poorly determined in the fit, we fix $\beta_{\rm m}=0.5$, the value derived for DLA host halos \citep{2012JCAP...11..059F,2018MNRAS.473.3019P}. Transmission biases are assumed to evolve with redshift as a power-law with exponent $\gamma_{\rm m}=1$, similar to the measured evolution of the CIV bias \citep{2018JCAP...05..029B}, but our results are not sensitive to this choice.

%--------------------------------------
\subsection{Other correlation terms}
%--------------------------------------
\label{subsec:otherterm}

The presence of HCDs in the absorption spectra modifies the expected correlation function. The flux transmission of spectra with identified DLAs are estimated by masking the strong absorption regions (transmission less than 20\%) and correcting the wings using a Voigt profile following the procedure of \citet{Lee13}. If this procedure worked perfectly, we would expect no strong modification of the power spectrum. However, it does not operate for HCDs below the nominal threshold of $\log N_{HI}\approx20$, and even above this threshold the detection efficiency depends on the signal-to-noise ratio of the spectrum. These imperfections modify the expected power spectrum.

We model the correlations due to absorption by unidentified HCD systems by adding to the power spectrum a term with the same form as the usual \lya~correlations (eqn. \ref{equation::pk_1}) but with $d_{\alpha}$ replaced by

\begin{equation}
  d_{\rm HCD}(\vec{k}) = b_{\rm HCD}^{\rm Ly\alpha-quasar}(z)\left(1 + \beta_{\rm HCD}\mu_{k}^{2}\right) F_{\rm HCD}(L_{\rm HCD}k_\parallel)
%  \exp{\left(-L_{\rm HCD}k_{\parallel}\right)}\ ,
\end{equation}
where the bias $b_{\rm HCD}^{\rm Ly\alpha-quasar}$ and the redshift-space distortion $\beta_{\rm HCD}$ are free parameters in the fit. The function $F_{\rm HCD}(L_{\rm HCD}k_\parallel)$ describes the suppression of power at large $k_\parallel$ due to unidentified HCDs of typical extent $L_{\rm HCD}$. The studies of mock data sets by \citet{2017A&A...603A..12B} tried several functional forms and $F=\sinc(L_{\rm HCD}k_\parallel)$ was adopted by them and by dMdB17, though other forms gave similar results. Following the more detailed studies of \citet{2018MNRAS.476.3716R}, we choose to use the form $F=\exp{\left(-L_{\rm HCD}k_{\parallel}\right)}$.

Our DLA-identification procedure requires their width
(wavelength interval for absorption greater than 20\% )
to be above $\sim2.0$~nm, corresponding to $\sim14~\hMpc$.
Following the study of \citet{2018MNRAS.476.3716R}, the corresponding unidentified HCD systems are well-modeled with
$L_{\rm HCD}=10\hMpc$ and we fix $L_{\rm HCD}$ to this value in the fits. We have verified that varying this parameter over the range $5<L_{\rm HCD}<15\hMpc$ does not change the fit position of the BAO peak. Due to degeneracies, we add a Gaussian prior on $\beta_{\rm HCD}$ of mean 0.5 and standard deviation 0.2.

The term in (\ref{equation::theoretical_cosmo}) representing
the transverse proximity effect takes the form
\citep{2013JCAP...05..018F}:
\begin{equation}
  \xi^{\rm TP} =
  \xi^{\rm TP}_{0}
  \left(
  \frac{1\,\hMpc}{r}
  \right)^{2}
  \exp(-r/\lambda_{\rm UV})\ .
\end{equation}
This form supposes isotropic emission from the quasars. We fix $\lambda_{\rm UV}=300~\hMpc$ \citep{2013ApJ...769..146R} and fit for the amplitude $\xi^{\rm TP}_{0}$.

In addition to accounting for asymmetries in the cross-correlation introduced by metal absorptions, continuum-fitting distortion and systematic redshift errors, the standard fit includes modeling of relativistic effects \citep{2014PhRvD..89h3535B}. The relativistic correction in (\ref{equation::theoretical_cosmo}) is the sum of two components describing a dipole and an octupole,
\begin{equation}
\xi^{\rm rel}(r,\mu)=A_{\rm rel1}\nu_1(r)L_1(\mu) + A_{\rm rel3}\nu_3(r)L_3(\mu)\ ,
\label{equation::rel}
\end{equation}
where $L_1$ and $L_3$ are the Legendre polynomial of degree 1 and 3 respectively, $A_{\rm rel1}$ and $A_{\rm rel3}$ are the amplitudes, and
\begin{equation}
\nu_\ell(r)=\frac{H_{0}}{c}\int kP_{\rm L}(k)j_\ell(kr)dk\ ,
\end{equation}
where $j_\ell$ is the spherical Bessel function. The relativistic dipole is expected to be the dominant contribution of odd-$\ell$ asymmetry and our standard fit therefore neglects the relativistic octupole ($A_{\rm rel3}=0$).

Dipole and octupole asymmetries also arise in the ``standard'' correlation function due to the evolution of the tracer biases and growth factor, as well as from the wide-angle correction  \citep{2014PhRvD..89h3535B}:
\begin{equation}
\xi^{\rm asy}(r,\mu)=\left(A_{\rm asy0}\eta_0(r)+A_{\rm asy2}\eta_2(r)\right)rL_1(\mu) + A_{\rm asy3}\eta_2(r)rL_3(\mu)\ ,
\label{equation::asy}
\end{equation}
where
\begin{equation}
\eta_\ell(r)=\frac{H_{0}}{c}\int k^2P_{\rm L}(k)j_\ell(kr)dk\ .
\end{equation}
Here, the two amplitudes $A_{\rm asy0}$ and $A_{\rm asy2}$ determine the dipole contribution, while $A_{\rm asy3}$ is the octupole amplitude. The $\xi^{\rm asy}$ term is neglected in the standard fit, but we check the robustness of the BAO measurement with respect to the odd-$\ell$ multipoles in Table~\ref{table::nonstandard}.

%--------------------------------------
\subsection{Broadband function}
%--------------------------------------
\label{subsec:broadband}

The optional $\xi^{\rm bb}$ term of (\ref{equation::xi_complete})
is a  ``broadband function'' that is
a slowly varying function of $(\rpar,\rperp)$:
\begin{equation}
\xi^{\rm bb}(r,\mu) = \sum^{i_{\rm max}}\limits_{i=i_{\rm min}} \sum^{j_{\rm max}}\limits_{j=_{\rm min}} a_{ij} \frac{L_{j}(\mu)}{r^{i}}\ ,
\label{equation::bb}
\end{equation}
where $L_{j}$  is the Legendre polynomial of degree $j$. Its purpose is to account for unknown physical, instrumental or analytical effects missing in the model that could potentially impact the BAO measurement. The standard fit features no broadband function. The result of adding a broadband function of the form $(i_{min},i_{max},j_{min},j_{max})=(0,2,0,6)$ is presented in Table~\ref{table::nonstandard}.

%===================================================
\section{Fits of the cross-correlation}
\label{section::fits}
%===================================================

\begin{figure*}[tb]
\centering
\includegraphics[width=\columnwidth]{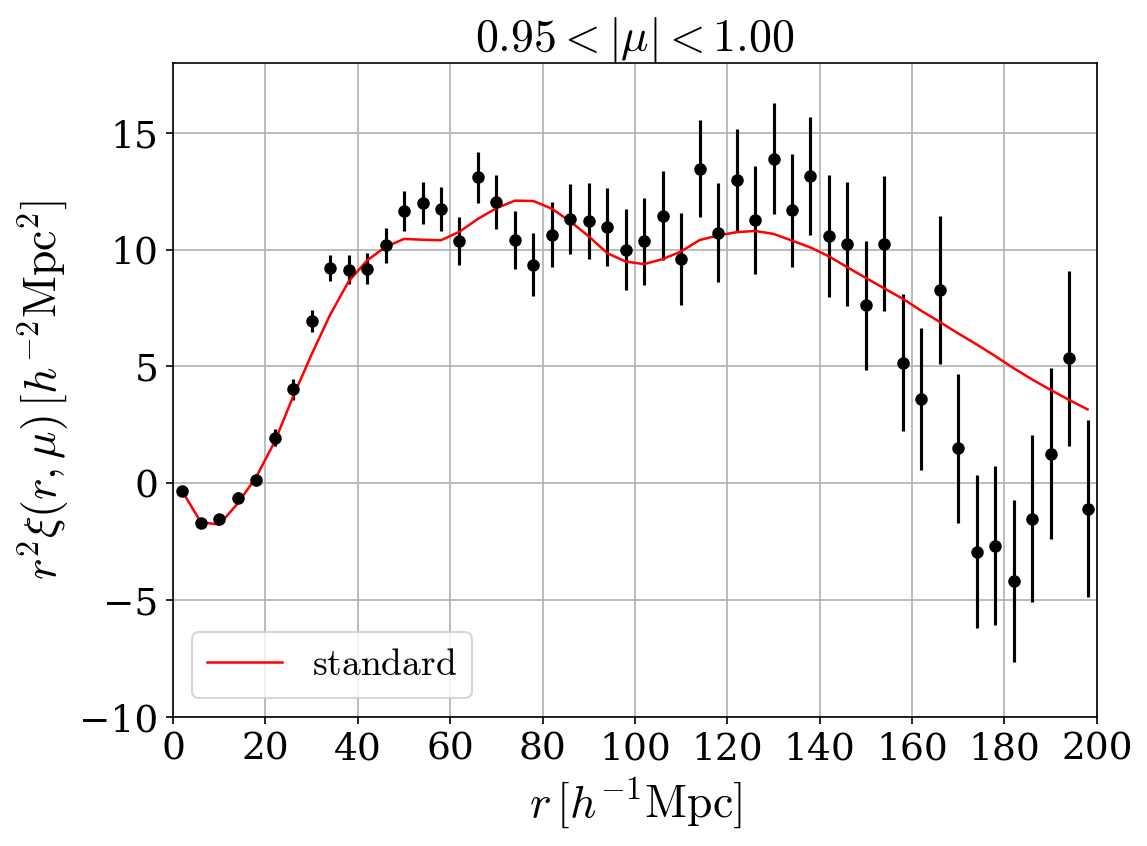}
\includegraphics[width=\columnwidth]{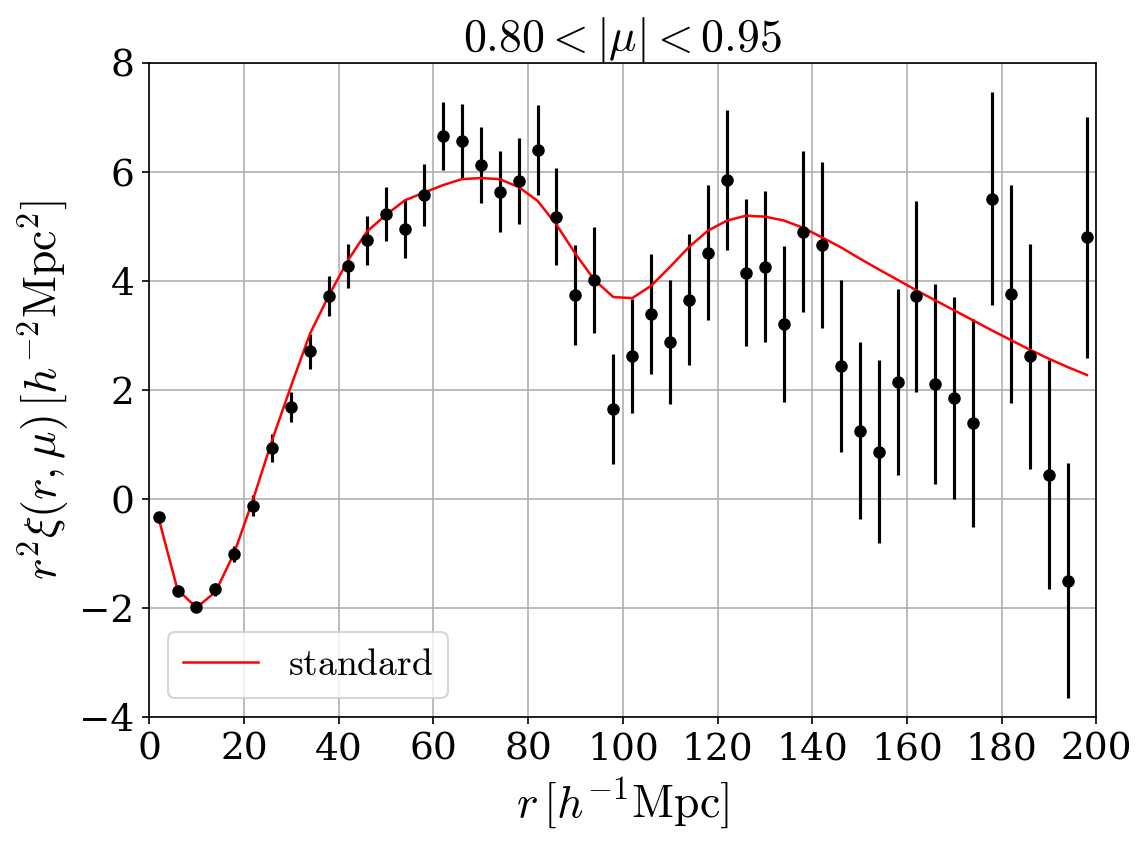}
\includegraphics[width=\columnwidth]{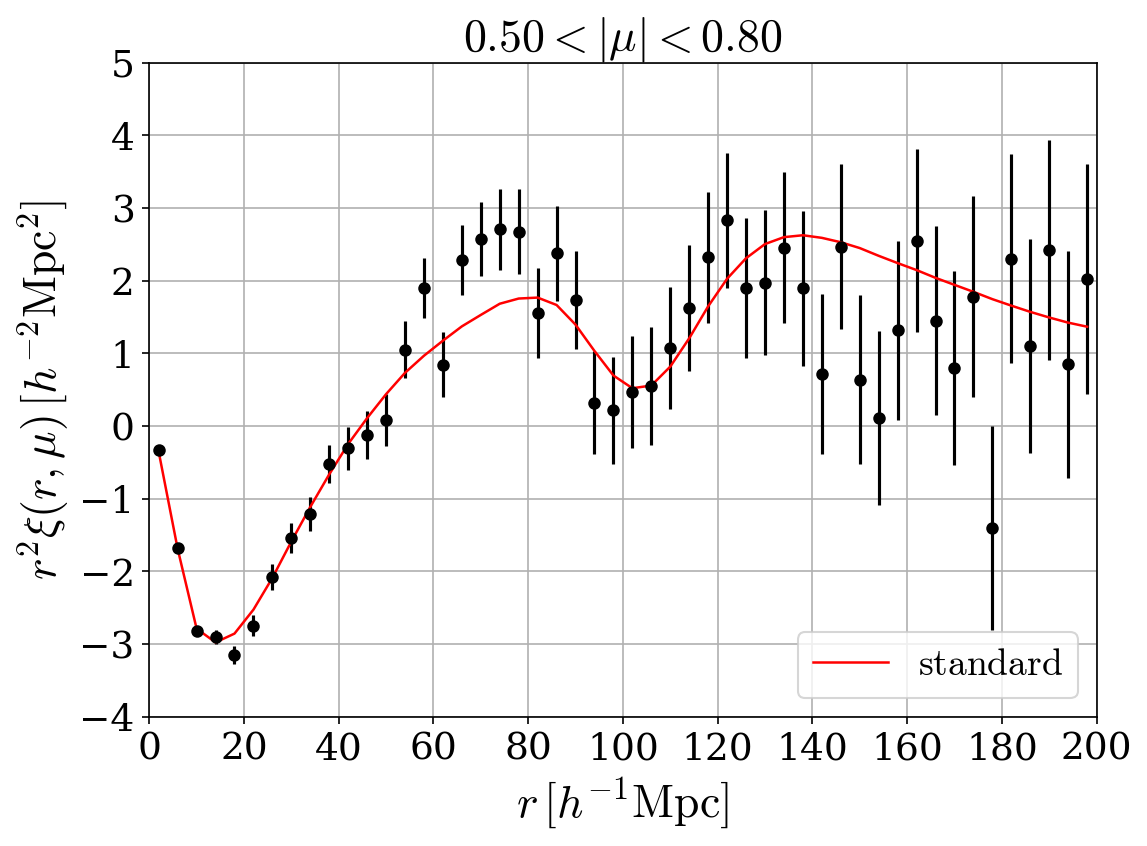}
\includegraphics[width=\columnwidth]{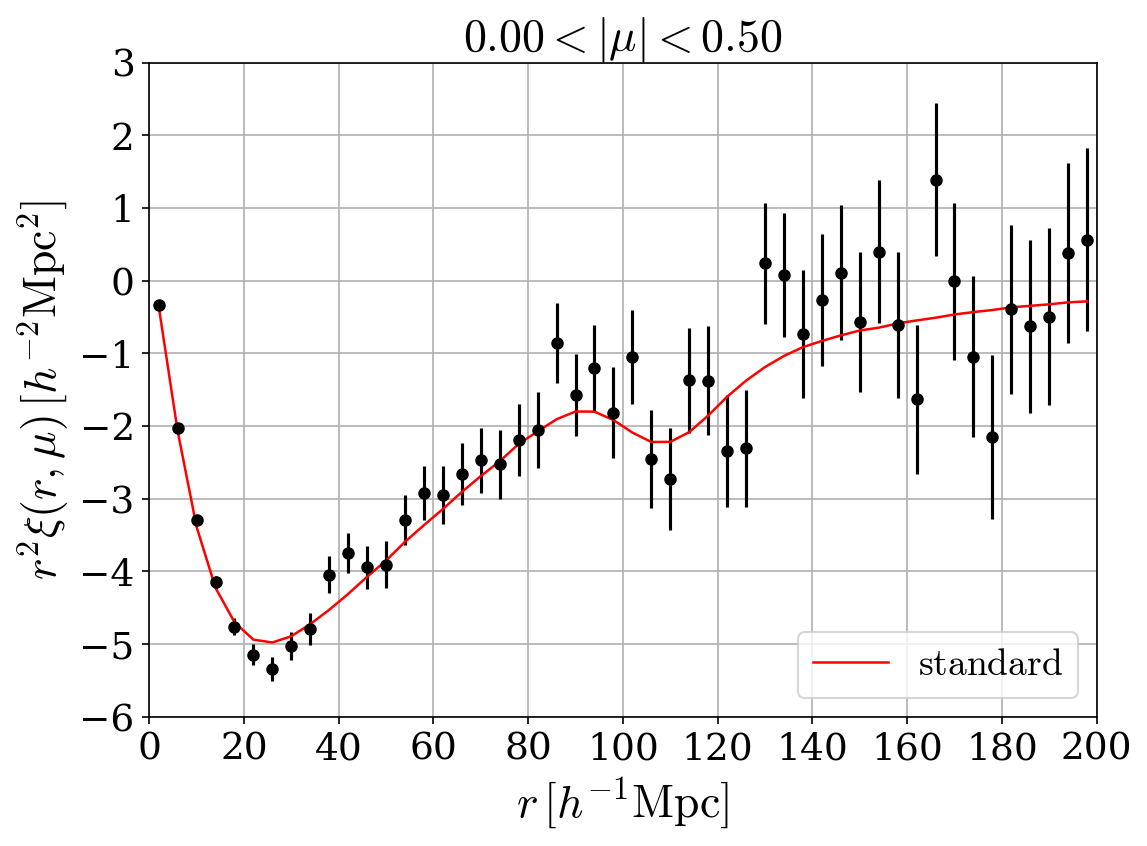}
\caption{Cross-correlation function averaged in four ranges of $\mu=r_{\parallel}/r$. The red curves show the best-fit model of the standard fit obtained for the fitting range $10<r<180~\hMpc$. The curves have been extrapolated outside this range.}
\label{figure::xi_wedges}
\end{figure*}

\begin{figure*}[tb]
\centering
\includegraphics[width=\columnwidth]{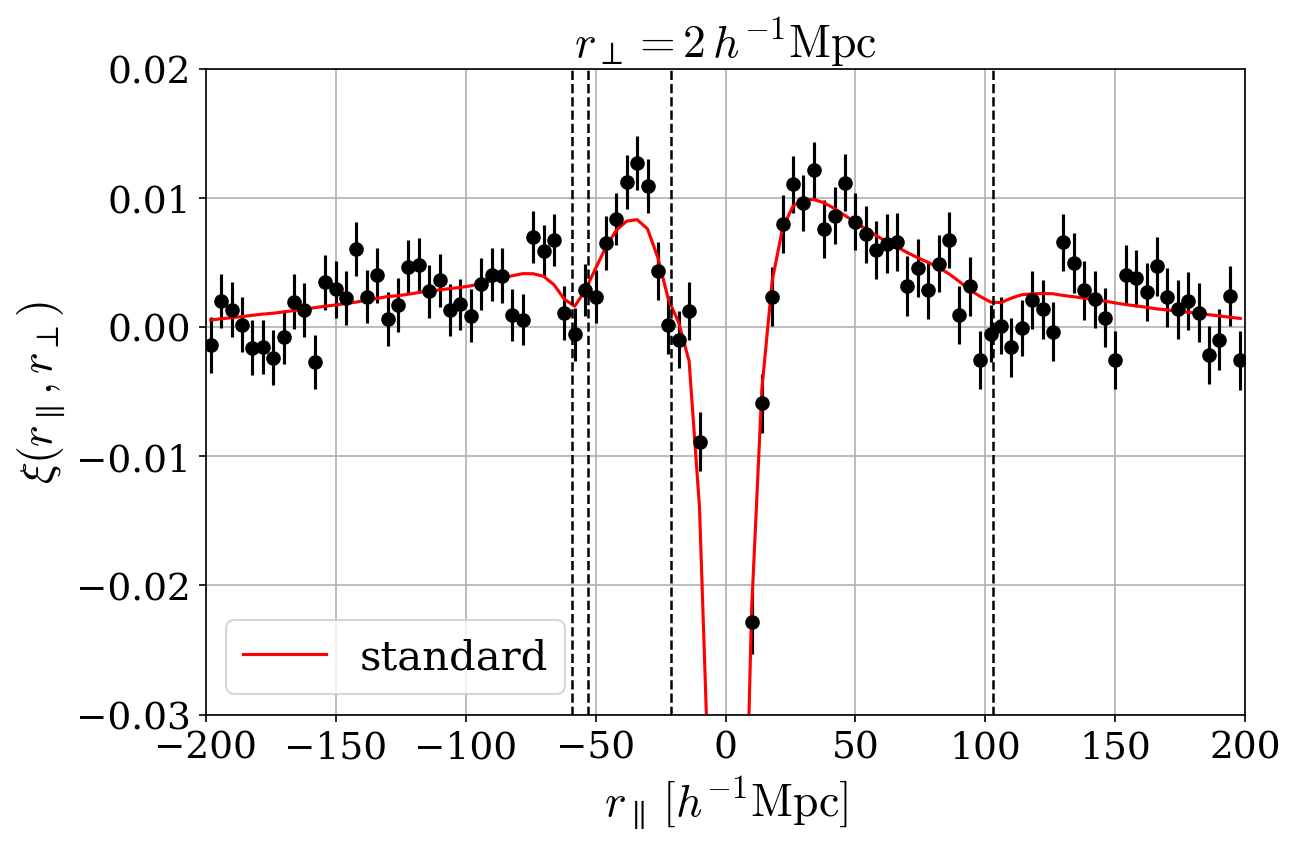}
\includegraphics[width=\columnwidth]{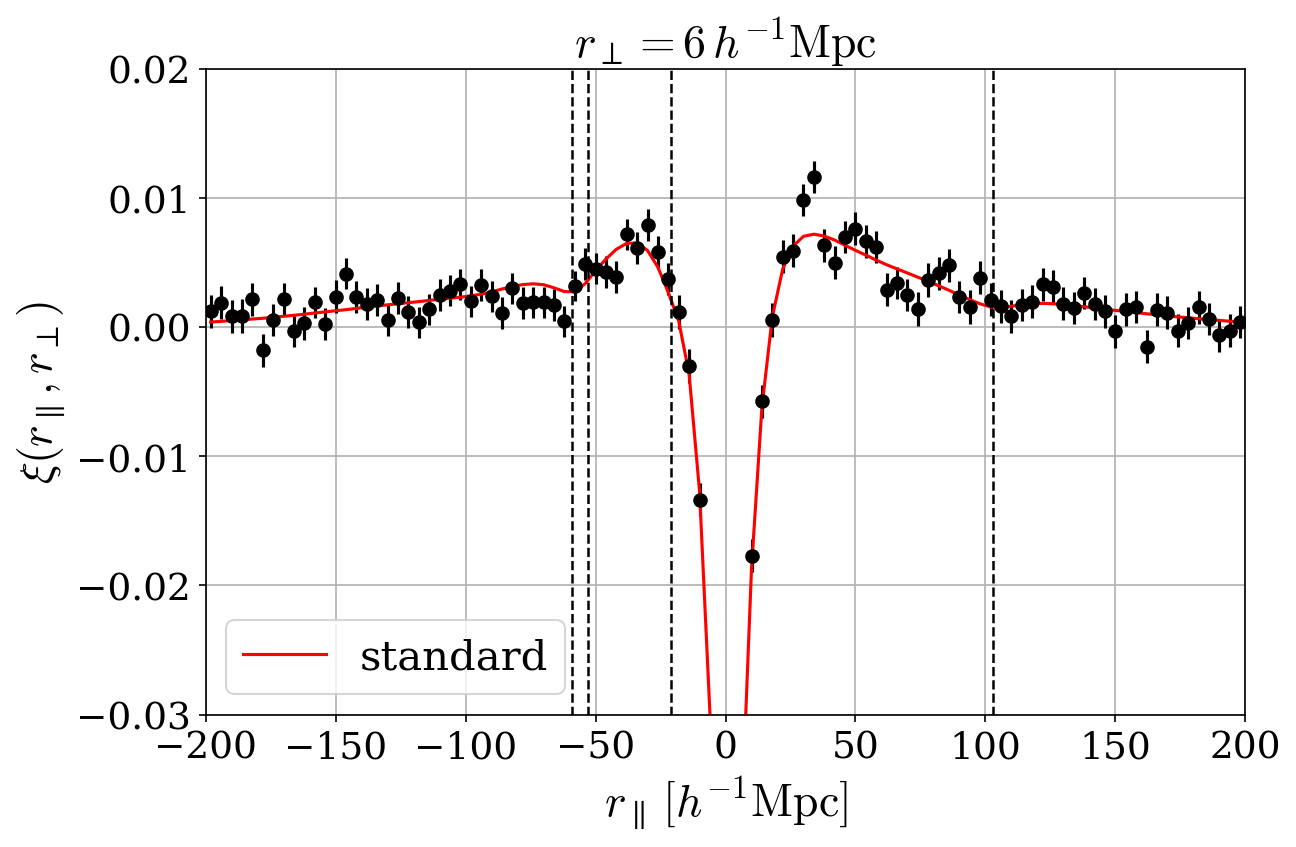}
\caption{Cross-correlation function as a function of $r_{\parallel}$ for two lowest values $r_{\perp}=[2,6]~\hMpc$. The red curves indicate the best-fit model of the standard fit obtained for the fitting range $10<r<180~\hMpc$. The curves have been extrapolated outside this range. The imprints of quasar-metal correlations are visible as peaks indicated by the dashed black lines at $r_{\parallel}\approx-21~\hMpc$ (SiIII(120.7)), $r_{\parallel}\approx-53~\hMpc$ (SiII(119.0)), $r_{\parallel}\approx-59~\hMpc$ (SiII(119.3)), and $r_{\parallel}\approx+103~\hMpc$ (SiII(126.0)).}
\label{figure::xi_slices}
\end{figure*}

\begin{table*}[tb]
\centering
\caption{Fit results for cross-correlation, auto-correlation of \citet{2019agathe}, and combined fit. The auto-correlation fit uses the combination Ly$\alpha$(Ly$\alpha$)xLy$\alpha$(Ly$\alpha$) + Ly$\alpha$(Ly$\alpha$)xLy$\alpha$(Ly$\beta$) described in \citet{2019agathe}. The fits are over the range $10<r<180~\hMpc$. Errors on BAO parameters correspond to $CL=68.27\%$, while the other parameters have errors corresponding to $\Delta\chi^{2}=1$. The parameter $\beta_{\rm q}$ is fixed for the cross-correlation fit. The bottom section of the table gives the minimum $\chi^2$, the number of data bins ($N_{\rm bin}$) and free parameters ($N_{\rm param}$) in the fit, the probability, the effective redshift, the correlation coefficient ($\rho$) for the BAO parameters, and the $\chi^2$ for the fit with the fixed fiducial BAO peak position.}
\label{table::bestfit}
\begin{tabular}{l r r r}
\hline
\hline
\noalign{\smallskip}
Parameter & Ly$\alpha$-quasar & Ly$\alpha$-Ly$\alpha$ & combined \\
\noalign{\smallskip}
\hline
\noalign{\smallskip}
$\alpha_{\parallel}$ & $1.076\pm0.042$ & $1.033\pm0.034$ & $1.049\pm0.026$ \\
$\alpha_{\perp}$ & $0.923\pm0.046$ & $0.953\pm0.048$ & $0.942\pm0.031$ \\
$\beta_{\alpha}$ & $2.28\pm0.31$ & $1.93\pm0.10$ & $1.99\pm0.10$ \\
$b_{\eta\alpha}$ & $-0.267\pm0.014$ & $-0.211\pm0.004$ & $-0.214\pm0.004$ \\
$\beta_{\rm q}$ & $0.257$ & & $0.209\pm0.006$ \\
$\sigma_{v}~[\hMpc]$ & $7.60\pm0.61$ & & $7.05\pm0.36$ \\
$\Delta r_{\parallel}~[\hMpc]$ & $-0.22\pm0.32$ & & $-0.17\pm0.28$ \\
$\xi^{\rm TP}_{0}$ & $0.276\pm0.158$ & & $0.477\pm0.112$ \\
$A_{\rm rel1}$ & $-13.5\pm5.8$ & & $-13.6\pm4.7$ \\
$\beta_{\rm HCD}$ & $0.500\pm0.200$ & $1.031\pm0.153$ & $0.972\pm0.150$ \\
$b_{\rm HCD}^{\rm Ly\alpha-quasar}$ & $-0.000\pm0.004$ & & $-0.000\pm0.004$ \\
$b_{\rm HCD}^{\rm Ly\alpha(Ly\alpha)-Ly\alpha(Ly\alpha)}$ & & $-0.051\pm0.004$ & $-0.052\pm0.004$ \\
$b_{\rm HCD}^{\rm Ly\alpha(Ly\alpha)-Ly\alpha(Ly\beta)}$ & & $-0.072\pm0.005$ & $-0.073\pm0.005$ \\
$10^{3}~b_{\rm SiII(119.0)}$ & $-5.7\pm2.4$ & $-5.0\pm1.0$ & $-4.3\pm0.9$ \\
$10^{3}~b_{\rm SiII(119.3)}$ & $-1.5\pm2.4$ & $-4.6\pm1.0$ & $-3.4\pm0.9$ \\
$10^{3}~b_{\rm SiIII(120.7)}$ & $-11.7\pm2.4$ & $-8.0\pm1.0$ & $-8.3\pm0.9$ \\
$10^{3}~b_{\rm SiII(126.0)}$ & $-2.2\pm1.7$ & $-2.2\pm1.3$ & $-1.9\pm0.9$ \\
$10^{3}~b_{\rm CIV(154.9)}$ & & $-16.3\pm8.8$ & $-16.8\pm9.0$ \\
\noalign{\smallskip}
\hline
\noalign{\smallskip}
$\chi_{\rm min}^{2}$ & $3231.61$ & $3258.91$ & $6499.31$ \\
$N_{\rm bin}$ & $3180$ & $3180$ & $6360$ \\
$N_{\rm param}$ & $14$ & $12$ & $18$ \\
probability & $0.20$ & $0.13$ & $0.08$ \\
$z_{\rm eff}$ & $2.35$ & $2.34$ & $2.34$ \\
$\rho(\alpha_{\parallel},\alpha_{\perp})$ & $-0.44$ & $-0.34$ & $-0.40$ \\
$\chi^{2}(\alpha_{\perp}=\alpha_{\parallel}=1)$ & $3235.79$ & $3260.54$ & $6504.30$ \\
\noalign{\smallskip}
\hline
\end{tabular}
\end{table*}

\begin{table}[tb]
\centering
\caption{Values of $\Delta\chi^{2}$ corresponding to $CL=(68.27,95.45\%)$. Values are derived from 10,000 Monte Carlo simulations of the correlation function that are fit using the model containing only Ly$\alpha$ absorption. Confidence levels are the fractions of the generated data sets that have best fits below the $\Delta\chi^{2}$ limit. The uncertainties are statistical and estimated using bootstrap.}
\label{table::confidence}
\begin{tabular}{l c c}
\hline
\hline
\noalign{\smallskip}
Parameter & $\Delta\chi^{2}$ $(68.27\%)$ & $\Delta\chi^{2}$ $(95.45\%)$ \\
\noalign{\smallskip}
\hline
\noalign{\smallskip}
Ly$\alpha$-quasar & & \\
$\alpha_{\parallel}$ & $1.15\pm0.02$ & $4.48\pm0.10$ \\
$\alpha_{\perp}$ & $1.15\pm0.02$ & $4.51\pm0.07$ \\
$(\alpha_{\parallel},\alpha_{\perp})$ & $2.51\pm0.03$ & $6.67\pm0.11$ \\
$F_{AP}$ & $1.13\pm0.02$ & $4.74\pm0.10$ \\
\noalign{\smallskip}
\hline
\noalign{\smallskip}
combined & & \\
$\alpha_{\parallel}$ & $1.08\pm0.02$ & $4.29\pm0.10$ \\
$\alpha_{\perp}$ & $1.08\pm0.02$ & $4.28\pm0.10$ \\
$(\alpha_{\parallel},\alpha_{\perp})$ & $2.47\pm0.03$ & $6.71\pm0.13$ \\
$F_{AP}$ & $1.11\pm0.02$ & $4.39\pm0.10$ \\
\noalign{\smallskip}
\hline
\end{tabular}
\end{table}

\begin{figure}[tb]
\centering
\includegraphics[width=\columnwidth]{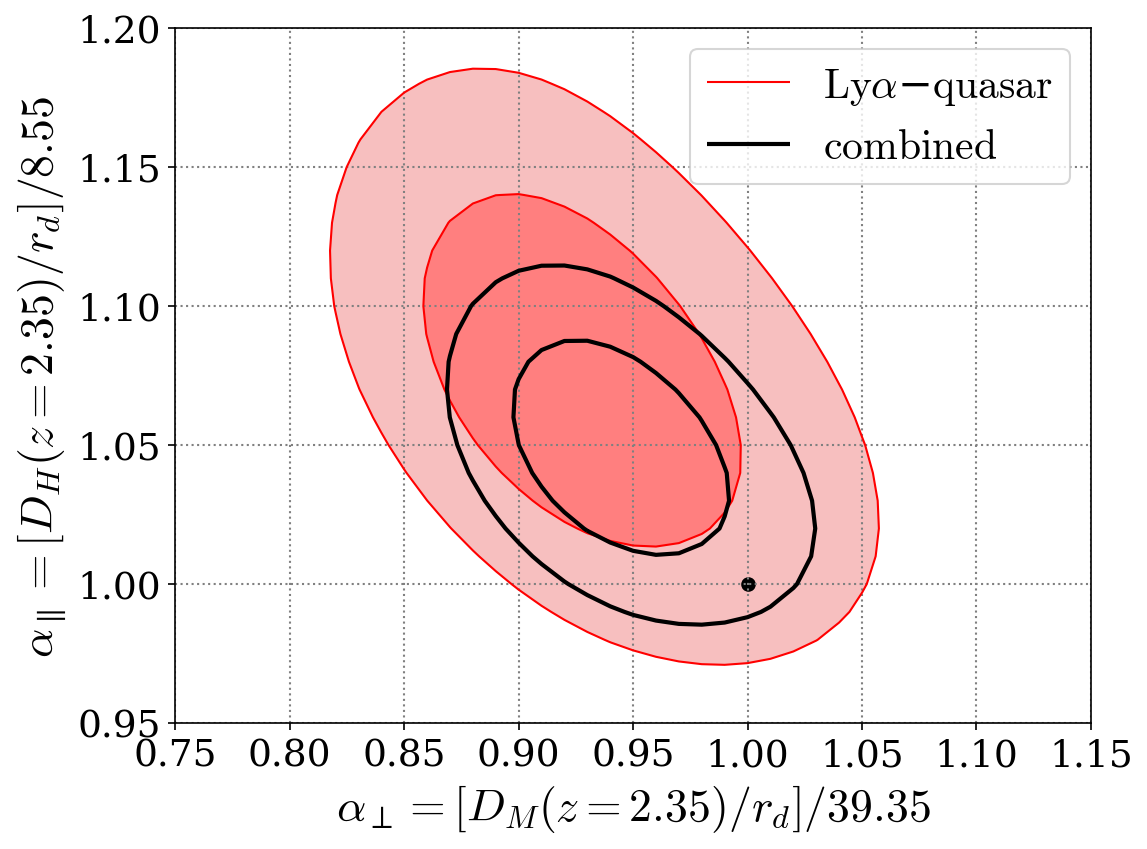}
\caption{Constraints on $(\alpha_{\parallel},\alpha_{\perp})$ for cross-correlation (red) and combination with auto-correlation (black). Contours correspond to confidence levels of $(68.27\%,95.45\%)$. The black point at $(\alpha_{\parallel},\alpha_{\perp})=(1,1)$ indicates the prediction of the Planck (2016) best-fit flat \lcdm~cosmology. The effective redshift of the combined fit is $z_{\rm eff}=2.34$ where the fiducial distance ratios are $(D_M/r_d,D_H/r_d)=(39.26,8.58)$.}
\label{figure::atap}
\end{figure}

\begin{figure}[tb]
\centering
\includegraphics[width=\columnwidth]{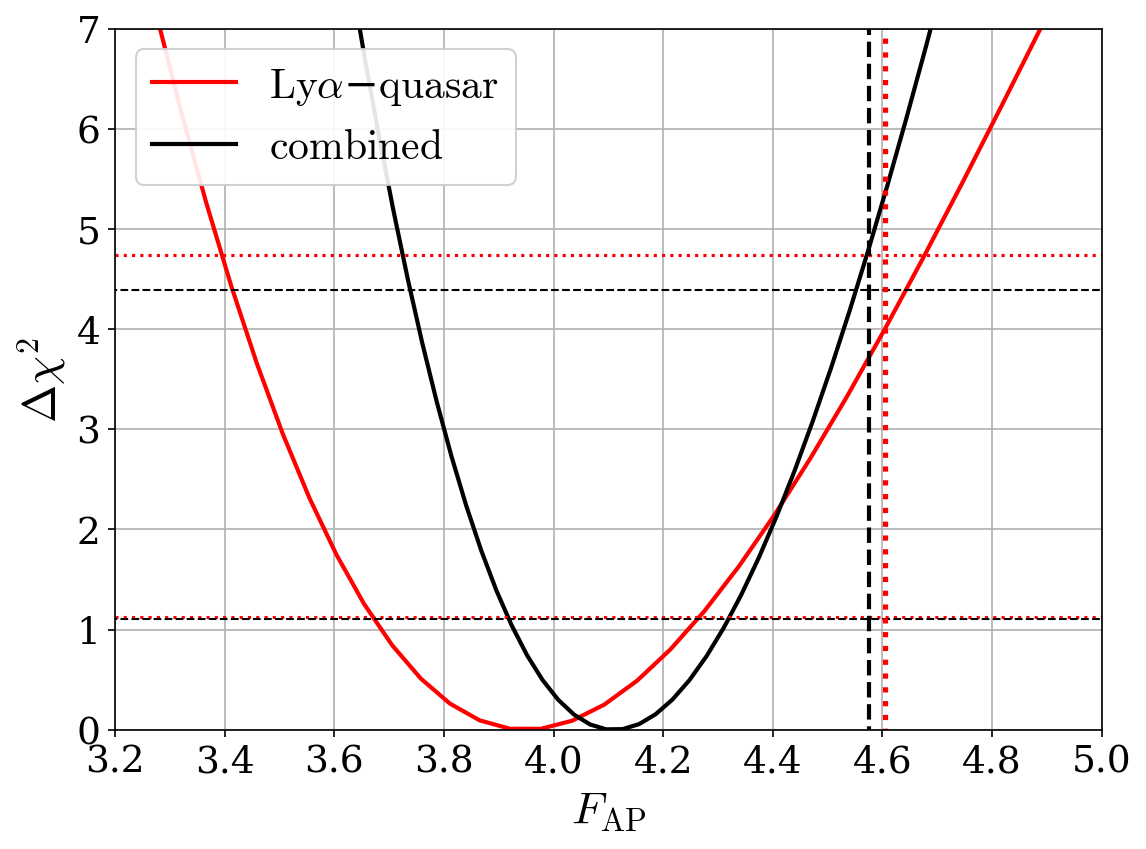}
\caption{Constraints on Alcock-Paczy\'nski parameter $F_{\rm AP}$ for cross-correlation (red) and combination with auto-correlation (black). Confidence levels of $(68.27\%,95.45\%)$ are indicated with the horizontal dotted lines for the cross-correlation and dashed lines for the combined fit. The prediction of the Planck (2016) best-fit flat \lcdm~cosmology is indicated with the vertical dotted line at \mbox{$F_{\rm AP}(z=2.35)=4.60$} for the cross-correlation and dashed line at \mbox{$F_{\rm AP}(z=2.34)=4.57$} for the combined fit.}
\label{figure::fap}
\end{figure}

Our standard fit of the cross-correlation function uses the
14 parameters in the first group of Table \ref{table::parameters}.
The fit includes 3180 data bins in the range $10<r<180~\hMpc$. The best-fit values are presented in the column ``Ly$\alpha$-quasar'' of Table~\ref{table::bestfit}. Figure~\ref{figure::xi_wedges} shows the best fit for four ranges of $\mu$ and Figure~\ref{figure::xi_slices} for the two lowest $\rperp$ bins.

Constraints on the BAO parameters $(\aperp,\apar)$ are presented in Fig.~\ref{figure::atap}. Following the method introduced and described in detail in dMdB17, we estimate the relation between $\Delta\chi^{2}=\chi^{2}-\chi_{\rm min}^{2}$ and confidence levels for the BAO parameters using a large number of simulated correlation functions generated from the best-fit model and the covariance matrix measured with the data. The results of the study, summarized in Table~\ref{table::confidence}, indicate that the (68.27,95.45\%) confidence levels for $(\aperp,\apar)$ correspond to $\Delta\chi^{2}=(2.51,6.67)$ (instead of the nominal values $\Delta\chi^{2}=(2.3,6.18)$). These levels are shown as contours in Fig.~\ref{figure::atap}. The best-fit values and confidence level (68.27,95.45\%) ranges are:
\begin{align}
 \aperp~=~ & 0.923~_{- 0.044 }^{+ 0.048 }\;_{- 0.087 }^{+ 0.105 }
        \label{equation::measure_aperp}\ ,\\[2pt]
\apar~=~ & 1.076~_{- 0.042 }^{+ 0.043 }\;_{- 0.085 }^{+ 0.088 }
        \label{equation::measure_apar}\ ,
\end{align}
corresponding to
\begin{align}
\frac{\DMm(z=2.35)}{r_{d}}~=~ & 36.3~_{- 1.7 }^{+ 1.9 }\;_{- 3.4 }^{+ 4.1 }
        \label{equation::measure_DArd}\ ,\\[2pt]
\frac{\DHh(z=2.35)}{r_{d}}~=~ & 9.20~_{- 0.36 }^{+ 0.37 }\;_{- 0.73 }^{+ 0.75 }
        \label{equation::measure_Dhrd}\ .
\end{align}
These results are consistent at 1.5 standard deviations with the prediction of the Planck (2016) best-fit flat $\Lambda$CDM model. Using a model without the BAO peak ($A_{\rm peak}=0$) degrades the quality of the fit by $\Delta\chi^2=22.48$.

Our BAO constraints can be compared with the DR12 measurement
of dMdB17 at a slightly higher redshift: $\DMm(2.40)/r_d=35.7\pm1.7$ and $\DHh(2.40)=9.01\pm0.36$ corresponding to $\aperp=0.898\pm0.042$ and $\apar=1.077\pm0.042$, relative to the same Planck model. The results (\ref{equation::measure_aperp}) and (\ref{equation::measure_apar}) thus represent a movement of $\sim0.3\sigma$ toward the Planck-inspired model through a shift in $\aperp$. As a cross-check of the results, we apply our analysis to the DR12 data set of dMdB17, without including the absorption in the Ly$\beta$ region. The best-fit values are $\aperp=0.889\pm0.040$ and $\apar=1.080\pm0.039$ (errors correspond to $\Delta\chi^{2}=1$), in good agreement with the measurement of dMdB17. This result indicates that the movement toward the fiducial model in DR14 is driven by the data.

Model predictions for $\DMm/r_d$ and $\DHh/r_d$ depend
both on pre-recombination physics, which determine $r_d$,
and on late-time physics, which determine $\DMm$ and $\DHh$.
Taking the ratio, yielding the Alcock-Paczy\'nski parameter $F_{AP}=D_{M}/D_{H}$ \citep{1979Natur.281..358A}, isolates the late-time effects which, in the \lcdm~model depend only on $(\Omega_{\rm m},\Omega_{\Lambda})$. We find
\begin{equation}
F_{\rm AP}(z=2.35)~=~3.95~_{- 0.28 }^{+ 0.32 }\;_{- 0.55 }^{+ 0.73 }
 \label{equation::measure_Fap}\ ,
\end{equation}
where the $\Delta\chi^2$ curve is shown in Fig.~\ref{figure::fap} and we have adopted that the (68.27,95.45\%) confidence levels correspond to $\Delta\chi^{2}=(1.13,4.74)$ (instead of the nominal values $\Delta\chi^{2}=(1,4)$). This result is 1.8 standard deviations from the prediction of the Planck-inspired model, $F_{\rm AP}(z=2.35)=4.60$.

The fit values of the \lya~bias parameters, $b_{\eta\alpha}=-0.267\pm0.014$ and $\beta_\alpha=2.28\pm0.31$ are consistent with those found by dMdB17, $b_{\eta\alpha}=-0.23\pm0.02$ and $\beta_\alpha=1.90\pm0.34$. These parameters can also be determined from the \lya~auto-correlation
and our value of $\beta_\alpha$ is consistent with that found with the auto-correlation function, $\beta_\alpha=1.93\pm0.10$ \citep{2019agathe}. However, these values are not in good agreement with the value $\beta_\alpha=1.656\pm0.086$ found earlier by \citet{2017A&A...603A..12B}. The auto- and cross-correlations values of $b_{\eta\alpha}$ also differ by $\sim20\%$:
$-0.267\pm0.014$ for the cross correlation and $-0.211\pm0.004$ for the auto-correlation. Furthermore, the bias parameters are not in good agreement with recent simulations \citep{2015JCAP...12..017A} which predict $\beta_\alpha\approx1.4$ and
$|b_{\eta\alpha}|$ in the range 0.14 to 0.20. Since our quoted uncertainties on the bias parameters (not on BAO parameters) come from approximating the likelihood as Gaussian, they might be underestimated in the presence of non-trivial correlations between the parameters. A dedicated study would be necessary to further investigate the consistency between the measured and predicted values. Fortunately, the bias parameters describe mostly the smooth
component of the correlation function and do not significantly
influence the BAO parameters $(\aperp,\apar)$, as indicated by the non-standard fits discussed below and summarized in Table \ref{table::nonstandard}.

The fit of the cross-correlation prefers a vanishing contribution from the quasar-HCD correlation term ($b_{\rm HCD}\approx0$). This preference is in contrast to the Ly$\alpha$ auto-correlation of \mbox{\citet{2019agathe}} where the HCD model is a crucial element to obtain a good fit (but does not affect the BAO peak position measurement). The best-fit radial coordinate shift $\Delta r_{\parallel}$ is consistent with zero systematic redshift error, but the parameter is strongly correlated with the amplitude of the relativistic dipole. Setting $A_{\rm rel1}=0$ in the fit yields $\Delta r_{\parallel}=-0.92\pm0.12$, in good agreement with the value reported in dMdB17. The best fit suggests marginal support for a non-zero value of $A_{\rm rel1}$, and the combined fit increases the significance of this result. However, even with sufficient statistical significance, its correlation with $\Delta r_{\parallel}$ (as well as other potential systematic errors on $A_{\rm rel1}$) prevents claims of a discovery of relativistic effects. The parameters $\sigma_{v}$ and $\xi_{0}^{\rm TP}$  have best-fit values in agreement with the result of dMdB17.

Among the metals, only SiIII(120.7) has a bias parameter significantly different from zero ($>4\sigma$) to show evidence for large-scale correlations with quasars. The imprints of the metal correlations are visible in the line-of-sight direction in Figure~\ref{figure::xi_slices}.

Besides the standard approach, we have also performed non-standard analyses, described in Appendix~\ref{section::nonstandard}, to search for unexpected systematic errors in the BAO peak-position measurement. The results for the non-standard fits of the cross-correlation are summarized in Table~\ref{table::nonstandard}. No significant changes of the best-fit values of $(\alpha_{\perp},\alpha_{\parallel})$ are observed. We have also divided the data to perform fits of the cross-correlation for a low- and a high-redshift bin as described in Appendix~\ref{section::zbins}. These fits are summarized in Table~\ref{table::zbins} and yield consistent best-fit BAO parameters for the two bins.

%===================================================
\section{Combination with the \lya~auto-correlation}
\label{section::combwithauto}
%===================================================

We combine our measurement of the Ly$\alpha$-quasar cross-correlation with the DR14 Ly$\alpha$ auto-correlation of \citet{2019agathe} by performing a combined fit of the correlation functions. For the auto-correlation, we use the combination Ly$\alpha$(Ly$\alpha$)xLy$\alpha$(Ly$\alpha$) + Ly$\alpha$(Ly$\alpha$)xLy$\alpha$(Ly$\beta$) described in \citet{2019agathe}. Because the covariance between the auto- and cross-correlation is sufficiently small to be ignored, as studied in \citet{2015A&A...574A..59D} and dMdB17, we treat their errors as independent. The combined fit uses the standard fit model of each analysis. In addition to the 14 free parameters in Table~\ref{table::parameters},  we let free the redshift-space distortion of quasars, $\beta_{\rm q}$, and the auto-correlation fit introduces three additional bias parameters ($b_{\rm CIV(154.9)}$, $b_{\rm HCD}^{\rm Ly\alpha(Ly\alpha)-Ly\alpha(Ly\alpha)}$, $b_{\rm HCD}^{\rm Ly\alpha(Ly\alpha)-Ly\alpha(Ly\beta)}$), for a total of 18 free parameters. The effective redshift of the combined fit is $z_{\rm eff}=2.34$.

The best-fit results are presented in the column ``combined'' of Table~\ref{table::bestfit}. Figure~\ref{figure::atap} displays the constraints on $(\aperp,\apar)$ from the combined measurement as black contours indicating the (68.27,95.45\%) confidence levels (corresponding to \mbox{$\Delta\chi^{2}=(2.47,6.71)$}; see Table~\ref{table::confidence}). The combined constraints on the BAO parameters are:
\begin{align}
 \aperp~=~ & 0.942~_{- 0.030 }^{+ 0.032 }\;_{- 0.059 }^{+ 0.067 }
        \label{equation::measure_aperp_comb}\ ,\\[2pt]
\apar~=~ & 1.049~_{- 0.025 }^{+ 0.026 }\;_{- 0.051 }^{+ 0.052 }
        \label{equation::measure_apar_comb}\ ,
\end{align}
corresponding to
\begin{align}
\frac{\DMm(z=2.34)}{r_{d}}~=~ & 37.0~_{- 1.2 }^{+ 1.3 }\;_{- 2.3 }^{+ 2.6 }
        \label{equation::measure_DArd_comb}\ ,\\[2pt]
\frac{\DHh(z=2.34)}{r_{d}}~=~ & 9.00~_{- 0.22 }^{+ 0.22 }\;_{- 0.43 }^{+ 0.45 }
        \label{equation::measure_Dhrd_comb}\ .
\end{align}
These results are within 1.7 standard deviations of the prediction of the Planck (2016) best-fit flat $\Lambda$CDM model. This movement of $\sim0.6\sigma$ toward the Planck prediction compared to the DR12 combined-fit result of dMdB17 is a consequence of the auto- and cross-correlation results individually moving toward the fiducial model.

Figure~\ref{figure::fap} shows the $\Delta\chi^{2}$ curve for the Alcock-Paczy\'nski parameter from the combined fit, for which the (68.27,95.45\%) confidence levels correspond to $\Delta\chi^{2}=(1.11,4.39)$. The combined constraint is
\begin{equation}
F_{\rm AP}(z=2.34)~=~4.11~_{- 0.19 }^{+ 0.21 }\;_{- 0.37 }^{+ 0.44 }
 \label{equation::measure_Fap_comb}\ ,
\end{equation}
within 2.1 standard deviations of the value \mbox{$F_{\rm AP}(z=2.34)=4.57$} expected in the Planck-inspired model.

%===================================================
\section{Implications for cosmological parameters}
\label{section::cosmology}
%===================================================

\begin{figure}[tb]
\centering
\includegraphics[width=\columnwidth]{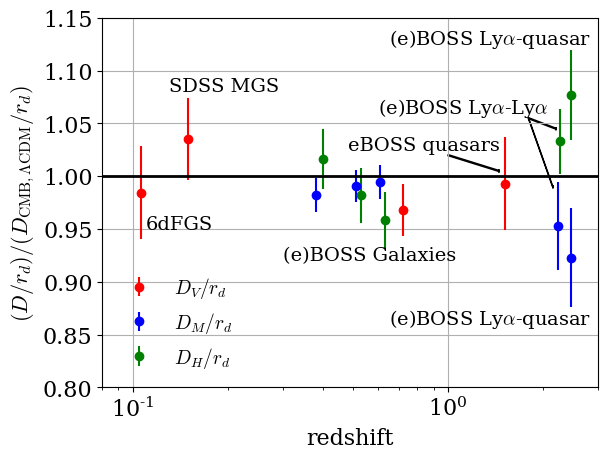}
\caption{Measurements of $D_M/r_d$, $D_H/r_d$ and $D_V/r_d$ at various
  redshifts:
  6dFGS \citep{2011MNRAS.416.3017B},
  SDSS MGS \citep{2015MNRAS.449..835R},
  BOSS galaxies \citep{2017MNRAS.470.2617A},
  eBOSS Galaxies \citep{2018ApJ...863..110B},
  eBOSS quasars \citep{2018MNRAS.473.4773A},
  eBOSS \lya-\lya~(de Sainte Agathe et al, 2019), and
  eBOSS \lya-quasars (this work).
  For clarity, the \lya-\lya~results at $z=2.34$ and
  the \lya-quasar results at $z=2.35$ have been separated slightly
  in the horizontal direction. Error bars represent $1\sigma$ uncertainties.
}
\label{figure::baoplot}
\end{figure}

\begin{figure}[tb]
\centering
\includegraphics[width=\columnwidth]{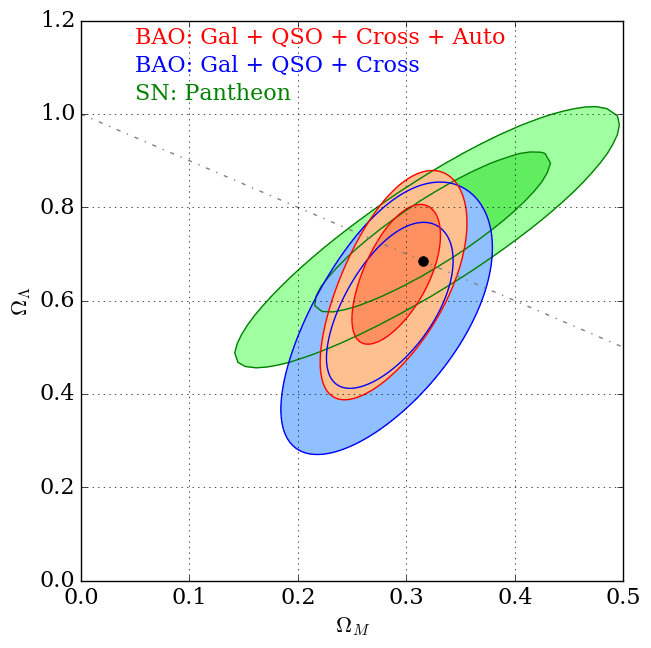}
\caption{One and two standard deviation constraints on
  $(\Omega_m,\Omega_\Lambda)$. The red contours use BAO measurements of  $D_M/r_d$ and $D_H/r_d$ of this work, of \citet{2019agathe} and \citet{2017MNRAS.470.2617A}, and the measurements of $D_V/r_d$ of \citet{2011MNRAS.416.3017B}, \citet{2015MNRAS.449..835R}, \citet{2018MNRAS.473.4773A} and \citet{2018ApJ...863..110B}. The blue contours do not use the \lya~auto-correlation measurement of \citet{2019agathe}. The green contours show the constraints from SN-Ia~Pantheon sample \citep{Scolnic18}. The black point indicates the values for the Planck (2016) best-fit flat \lcdm~cosmology.}
\label{figure::omol}
\end{figure}

\begin{figure}[tb]
\centering
\includegraphics[width=\columnwidth]{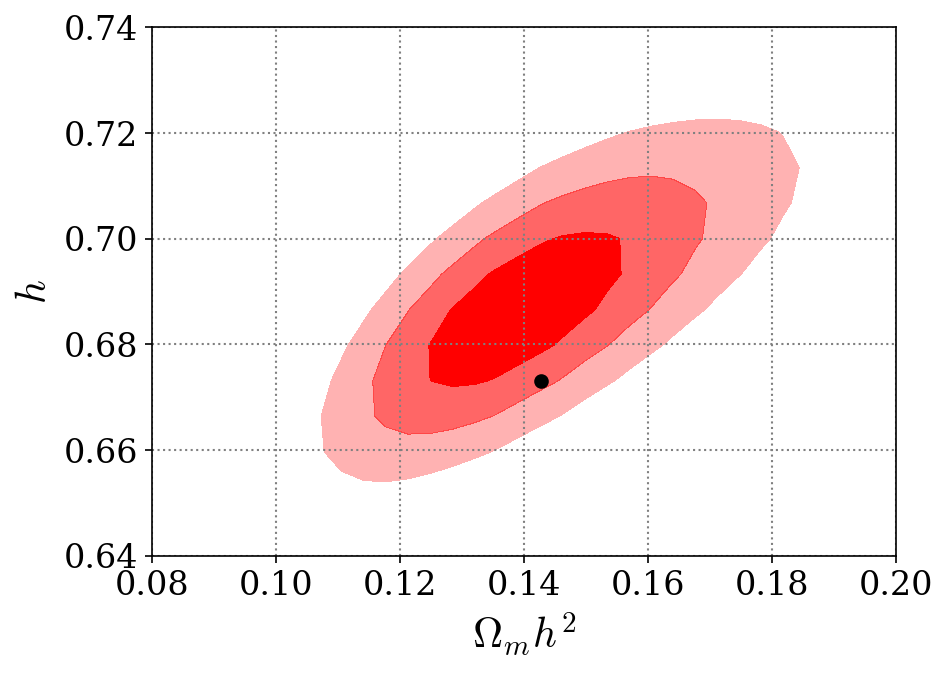}
\caption{One and two standard deviation constraints on $H_0$ and $\Omega_mh^2$ derived from BAO data used in Fig.~\ref{figure::omol} and from Big-Bang Nucleosynthesis. This figure assumes a flat universe and a Gaussian prior $100\Omega_bh^2=2.260\pm0.034$ derived from the deuterium
abundance measurement of \citet{2018ApJ...855..102C}.}
\label{figure::h0bbn}
\end{figure}

The combined-fit measurement of $(D_M/r_d,D_H/r_d)$ at \mbox{$z=2.34$} presented here is within 1.7 standard deviations of the predictions of the flat \lcdm~model favored by the measurement of CMB anisotropies \citep{2016A&A...594A..13P}. This result thus does not constitute statistically significant evidence for new physics or unidentified systematic errors in the measurement. Figure~\ref{figure::baoplot} illustrates the agreement with the Planck prediction for the ensemble of BAO measurements.

Independent of CMB data and without assuming flatness, the BAO data by themselves constrain the parameters $(\Omega_m,\Omega_\Lambda,H_0r_d)$ of the (o)\lcdm~model. Using the combined fit (eqns. \ref{equation::measure_DArd_comb} and \ref{equation::measure_Dhrd_comb}), the galaxy data of
\citet{2017MNRAS.470.2617A}, \citet{2011MNRAS.416.3017B}
\citet{2015MNRAS.449..835R} and \citet{2018ApJ...863..110B}
and the quasar data of \citet{2018MNRAS.473.4773A} yields
\begin{equation}
\Omega_M=0.293 \pm 0.027 \hspace*{10mm} \Omega_\Lambda=0.675 \pm 0.099 
\end{equation}
corresponding to $\Omega_k=0.032 \pm 0.117$.
The best fit gives $(c/H_0)/r_d=29.78 \pm0.56$ corresponding to
$hr_d=(0.683\pm 0.013)\times147.33$~Mpc. The CMB inspired flat~\lcdm~model  has $\chi^2=13.76$ for 12 degrees of freedom and is within one standard deviation of the best fit, as illustrated in Figure~\ref{figure::omol}.

A value of $H_0$ can be obtained either by using the CMB measurement of $r_d$ or by using the Primordial-Nucleosynethsis value of $\Omega_bh^2$ to constrain $r_d$. Adopting the value \mbox{$100\Omega_bh^2=2.260\pm0.034$} derived from the deuterium abundance measurement of \citet{2018ApJ...855..102C} and assuming flat \lcdm, we derive the constraints on ($H_0,\Omega_mh^2$) shown in Figure~\ref{figure::h0bbn} with
\begin{equation}
h=0.686\pm0.010
\end{equation}
or $h<0.706$ at 95\% C.L. The limit degrades to $h<0.724$ (95\% C.L.) if one adopts a more conservative uncertainty on the baryon density: $100\Omega_bh^2=2.26\pm0.20$. Nevertheless, as previously noted \citep{2015PhRvD..92l3516A,2018ApJ...853..119A}, the combination of BAO and nucleosynthsis provides a CMB-free confirmation of the tension with the distance-ladder determinations of $H_0$ \citep{2016ApJ...826...56R,2018ApJ...855..136R,2018ApJ...861..126R}.

%===================================================
\section{Conclusions}
\label{section::conclusions}
%===================================================

Using the entirety of BOSS and the first two years of eBOSS observations from SDSS DR14, this paper has presented a measurement of the cross-correlation of quasars and the \lya~flux transmission at redshift 2.35. In addition to the new and reobserved quasars provided in DR14, we have improved statistics further by extending the Ly$\alpha$ forest to include Ly$\alpha$ absorption in the Ly$\beta$ region of the spectra.

The position of the BAO peak is $1.5\sigma$ from the flat \lcdm~model favored by CMB anisotropy measurements \citep{2016A&A...594A..13P}. We emphasize that the measured peak position shows no significant variation when adding astrophysical elements to the fit model. The basic Ly$\alpha$-only model on its own provides an accurate determination of the peak position, while still yielding an acceptable fit to the data. Compared to the BAO measurement for the DR12 data set reported by dMdB17, our result represents a movement of $\sim0.3\sigma$ toward the Planck-cosmology prediction through a shift in the transverse BAO parameter $\aperp$. This change is driven by the data and not by differences in the analyses. The inclusion of Ly$\alpha$ absorption in the Ly$\beta$ region has no impact on the best-fit value of $\aperp$. Combined with the \lya-flux-transmission auto-correlation measurement presented in a companion paper \citep{2019agathe}, the BAO peak at $z=2.34$ is $1.7\sigma$ from the expected value.

The ensemble of BAO measurements is in good agreement with the CMB-inspired flat \lcdm~model. By themselves, the BAO data provide a good confirmation of this model. The use of SNIa to measure cosmological distances \citep{Scolnic18} provides independent measurements of the model parameters. As can be seen in Fig.~\ref{figure::omol} they are in agreement with the BAO measurements.

The BAO measurements presented here will be improved by the upcoming DESI \citep{2016arXiv161100036D} and WEAVE-QSO \citep{2016sf2a.conf..259P} projects both by increasing the number of quasars and improving the spectral resolution.

The best-fit results and the $\chi^{2}$ scans for the cross-correlation by itself and the combination with the auto-correlation are publicly available.\footnote{\url{https://github.com/igmhub/picca/tree/master/data/Blomqvistetal2019}}

\begin{acknowledgements}

We thank Pasquier Noterdaeme for providing the DLA catalog for eBOSS DR14 quasars.

This work was supported by the A*MIDEX project (ANR-11-IDEX-0001-02) funded by the “Investissements d’Avenir” French Government program, managed by the French National Research Agency (ANR), and by ANR under contract ANR-14-ACHN-0021.

Funding for the Sloan Digital Sky Survey IV has been provided by the Alfred P. Sloan Foundation, the U.S. Department of Energy Office of Science, and the Participating Institutions. SDSS acknowledges
support and resources from the Center for High-Performance Computing at the University of Utah. The SDSS web site is \url{http://www.sdss.org/}.

SDSS is managed by the Astrophysical Research Consortium for the Participating Institutions of the SDSS Collaboration including the Brazilian Participation Group, the Carnegie Institution for Science, Carnegie Mellon University, the Chilean Participation Group, the French Participation Group, Harvard-Smithsonian Center for Astrophysics, Instituto de Astrofísica de Canarias, The Johns Hopkins University, Kavli Institute for the Physics and Mathematics of the Universe (IPMU) / University of Tokyo, Lawrence Berkeley National Laboratory, Leibniz Institut für Astrophysik Potsdam (AIP), Max-Planck-Institut für Astronomie (MPIA Heidelberg), Max-Planck-Institut für Astrophysik (MPA Garching), Max-Planck-Institut für Extraterrestrische Physik (MPE), National Astronomical Observatories of China, New Mexico State University, New York University, University of Notre Dame, Observatório Nacional / MCTI, The Ohio State University, Pennsylvania State University, Shanghai Astronomical Observatory, United Kingdom Participation Group, Universidad Nacional Autónoma de México, University of Arizona, University of Colorado Boulder, University of Oxford, University of Portsmouth, University of Utah, University of Virginia, University of Washington, University of Wisconsin, Vanderbilt University, and Yale University.

A.F.R. was supported by an STFC Ernest Rutherford Fellowship, grant reference ST/N003853/1, and by STFC Consolidated Grant no ST/R000476/1.

\end{acknowledgements}

\bibliographystyle{aa}
\bibliography{qsolyaDR14}

\appendix

%===================================================
\section{Non-standard fits of the cross-correlation}
\label{section::nonstandard}
%===================================================

\begin{table*}[tb]
\centering
\caption{Results of non-standard fits. The first group presents results of successively adding complications from physical effects to the basic Ly$\alpha$-only model. These complications are: metals, absorption by high-column density systems, the transverse proximity effect, and the relativistic dipole, corresponding to the standard fit from column 1 of Table~\ref{table::bestfit}. The second group presents fits which include fluctuations of the UV background radiation, the odd multipoles $\ell=(1,3)$ or the broadband function (for this group we set $\xi^{\rm qHCD}=0$). The last group presents fits for non-standard data samples: no absorption in the Ly$\beta$ region or no correction of DLAs in the spectra. The fit is over the range $10<r<180~\hMpc$. Errors correspond to $\Delta\chi^{2}=1$.}
\label{table::nonstandard}
\begin{tabular}{l l l l l l}
\hline
\hline
\noalign{\smallskip}
Analysis & $\apar$ & $\aperp$ & $\beta_{\alpha}$ & $b_{\eta\alpha}$ & $\chiSquareMin/DOF, \,\, \mathrm{probability}$  \\
\noalign{\smallskip}
\hline
\noalign{\smallskip}
Ly$\alpha$ & $ 1.073\pm0.041 $ & $ 0.925\pm0.045 $ & $ 2.75\pm0.21 $ & $ -0.285\pm0.012 $ & $ 3268.55 / (3180-~~~6),  p = 0.12 $ \\
+ metals & $ 1.074\pm0.041 $ & $ 0.921\pm0.045 $ & $ 2.76\pm0.22 $ & $ -0.281\pm0.012 $ & $ 3239.52 / (3180-~10),  p = 0.19 $ \\
+ HCD & $ 1.074\pm0.041 $ & $ 0.921\pm0.045 $ & $ 2.76\pm0.22 $ & $ -0.281\pm0.017 $ & $ 3239.52 / (3180-~12),  p = 0.18 $ \\
+ TP & $ 1.075\pm0.040 $ & $ 0.923\pm0.043 $ & $ 2.31\pm0.30 $ & $ -0.269\pm0.014 $ & $ 3236.62 / (3180-~13),  p = 0.19 $ \\
+ rel1 & $ 1.076\pm0.040 $ & $ 0.923\pm0.043 $ & $ 2.28\pm0.31 $ & $ -0.267\pm0.014 $ & $ 3231.61 / (3180-~14),  p = 0.20 $ \\
\noalign{\smallskip}
\hline
\noalign{\smallskip}
UV & $ 1.077\pm0.040 $ & $ 0.923\pm0.043 $ & $ 2.34\pm0.32 $ & $ -0.274\pm0.020 $ & $ 3231.30 / (3180-~13),  p = 0.21 $ \\
odd-$\ell$ & $ 1.074\pm0.040 $ & $ 0.927\pm0.045 $ & $ 2.33\pm0.32 $ & $ -0.267\pm0.014 $ & $ 3223.25 / (3180-~16),  p = 0.23 $ \\
BB (0,2,0,6) & $ 1.083\pm0.039 $ & $ 0.921\pm0.043 $ & $ 2.53\pm0.46 $ & $ -0.280\pm0.022 $ & $ 3223.75 / (3180-~24),  p = 0.20 $ \\
\noalign{\smallskip}
\hline
\noalign{\smallskip}
no Ly$\beta$ & $ 1.084\pm0.040 $ & $ 0.921\pm0.042 $ & $ 2.33\pm0.32 $ & $ -0.272\pm0.014 $ & $ 3231.05 / (3180-~14),  p = 0.21 $ \\
keep DLAs & $ 1.071\pm0.042 $ & $ 0.929\pm0.049 $ & $ 2.08\pm0.27 $ & $ -0.279\pm0.016 $ & $ 3217.64 / (3180-~14),  p = 0.26 $ \\
\noalign{\smallskip}
\hline
\end{tabular}
\end{table*}

\begin{figure*}
\centering
\includegraphics[width=\columnwidth]{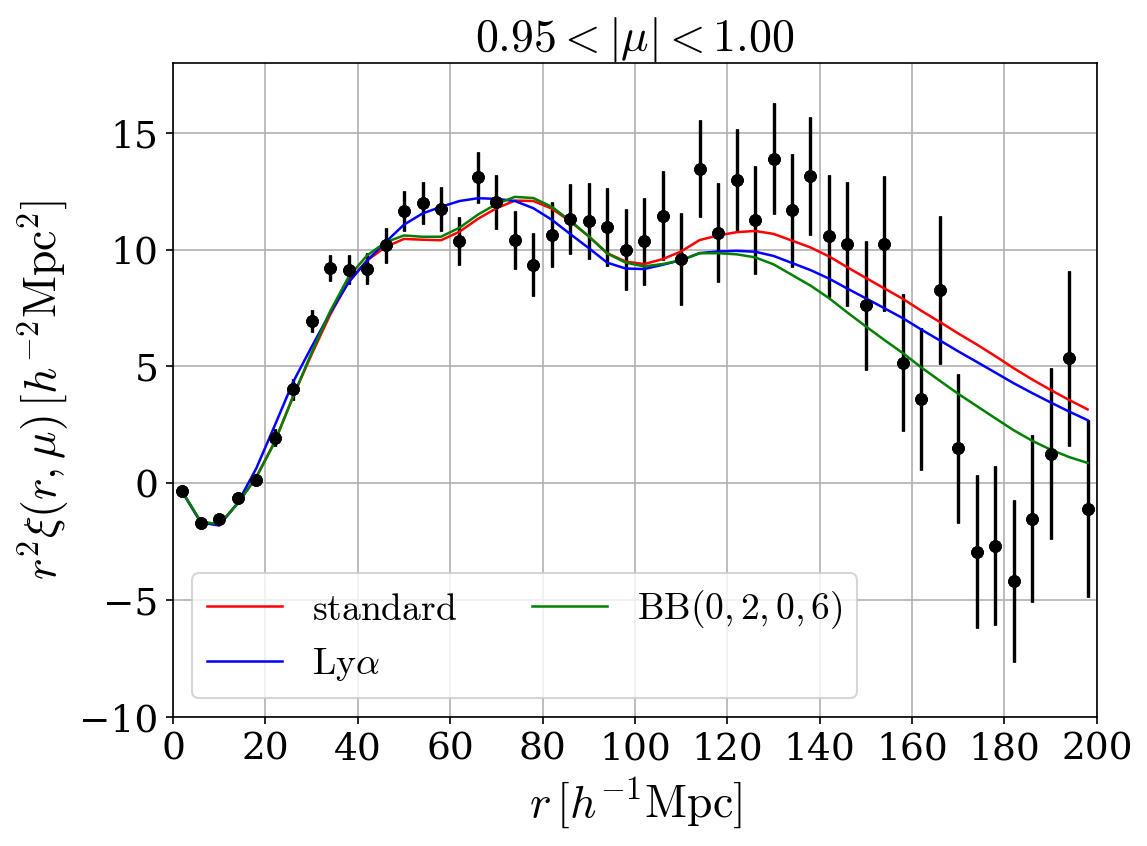}
\includegraphics[width=\columnwidth]{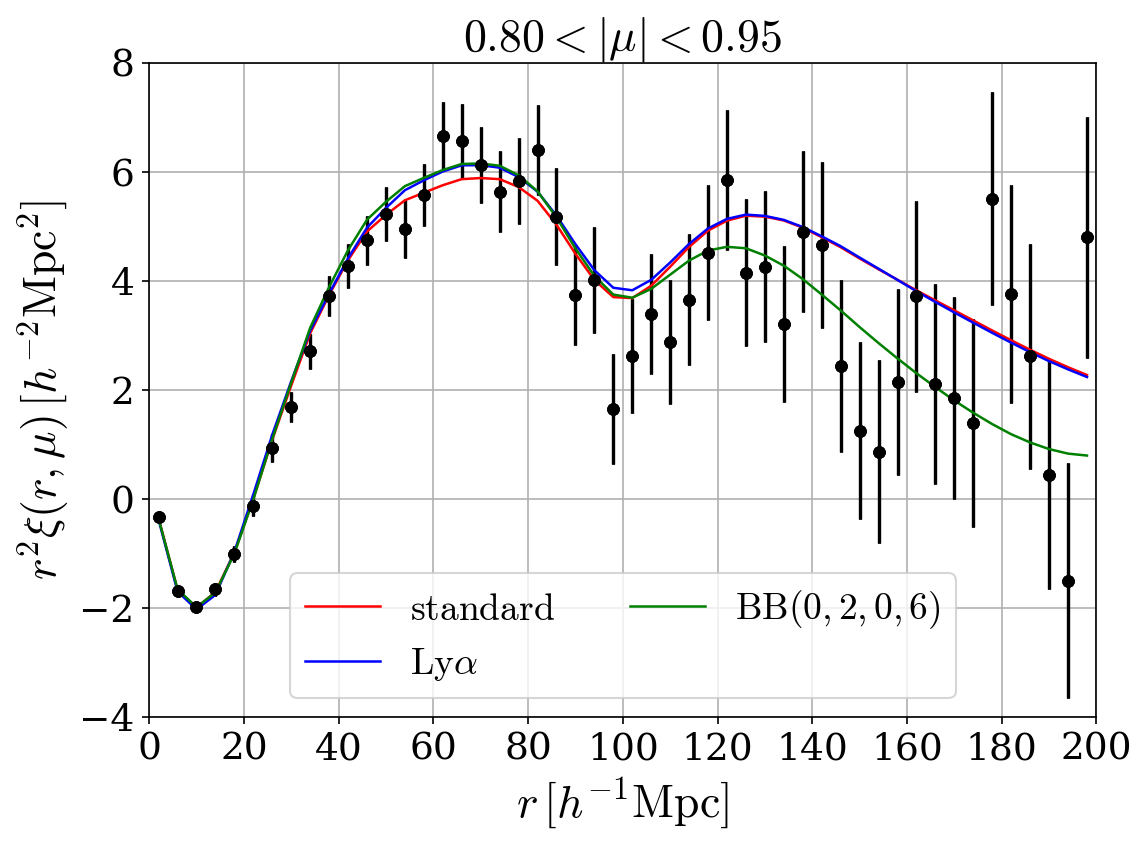}
\includegraphics[width=\columnwidth]{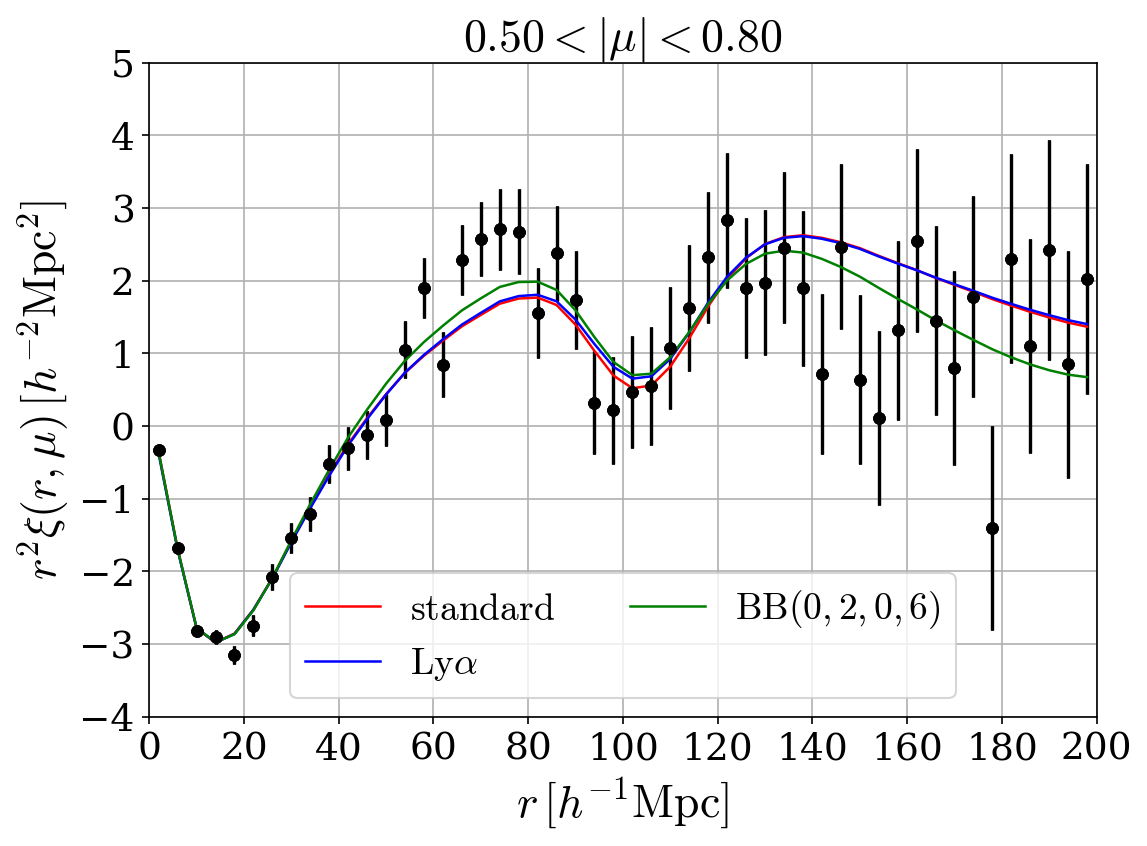}
\includegraphics[width=\columnwidth]{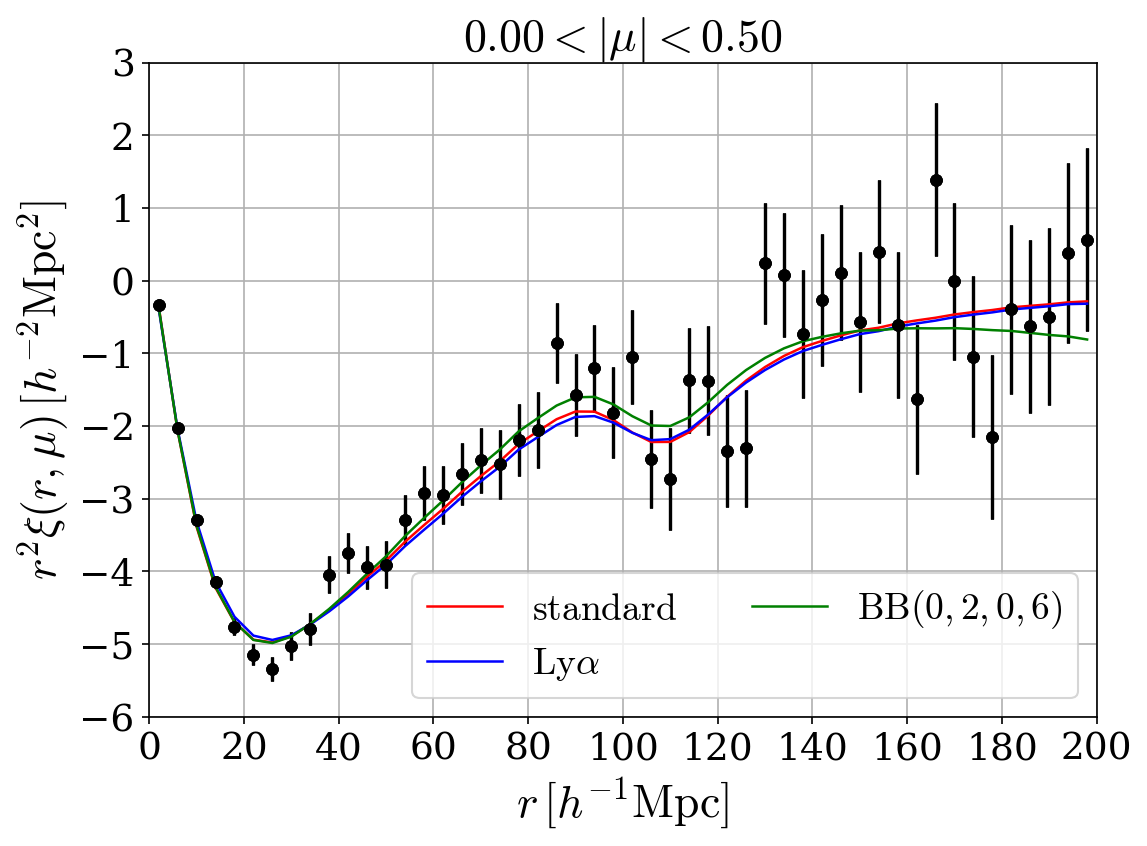}
\caption{Same as Figure~\ref{figure::xi_wedges} but showing three models fit to data. Red curves indicate the standard fit, blue curves the basic Ly$\alpha$-only model, and green curves the standard fit (with $\xi^{\rm qHCD}=0$) with the addition of the broadband function (equation~\ref{equation::bb}) of the form $(i_{min},i_{max},j_{min},j_{max})=(0,2,0,6)$. The curves have been extrapolated outside the fitting range $10<r<180~\hMpc$.}
\label{figure::xi_nonstd_wedges}
\end{figure*}

The results of performing non-standard fits of the cross-correlation are summarized in Table~\ref{table::nonstandard}. The first group reports results obtained by successively adding elements to the model, starting with a model with only the standard \lya~correlation function and ending with the complete model of Table~\ref{table::bestfit}. Adding elements does not significantly change the best-fit values of $(\aperp,\apar)$ while gradually improving the quality of the fit.

The second group of fits in Table~\ref{table::nonstandard} adds elements to the standard fit in the form of fluctuations of the UV background radiation (eqn~\ref{equation::uv}), the $A_{\rm rel3}$ term (eqn~\ref{equation::rel}) and the other odd-$\ell$ terms from equation~\ref{equation::asy}, or the broadband function (eqn~\ref{equation::bb}). For these fits, we set $\xi^{\rm qHCD}=0$ to facilitate the parameter error estimation with no impact on the best fits. No significant changes of the BAO parameters are observed.

The third group of fits concern non-standard data samples that either omits the correlation pairs from the Ly$\beta$ region (``no Ly$\beta$") or leaves the DLAs uncorrected in the spectra (``keep DLAs"). Even for these modified data samples the best-fit values of $(\aperp,\apar)$ do not deviate significantly from those of our standard analysis.

Figure~\ref{figure::xi_nonstd_wedges} shows the measured cross-correlation for four ranges of $\mu$ and three of the fits listed in Table~\ref{table::nonstandard}: the standard fit used to measure the BAO parameters, the basic Ly$\alpha$-only fit, and the fit with the broadband function.

%===================================================
\section{Redshift split}
\label{section::zbins}
%===================================================

\begin{table*}
\centering
\caption{Fit results for two redshift bins. The effective redshifts are $z_{\rm eff}=2.21$ ($z_{m}<2.48$) and $z_{\rm eff}=2.58$ ($z_{m}>2.48$). The fit is over the range $10<r<180~\hMpc$. Errors correspond to $\Delta\chi^{2}=1$.}
\label{table::zbins}
\begin{tabular}{l l l l l l}
\hline
\hline
\noalign{\smallskip}
Analysis & $\apar$ & $\aperp$ & $\beta_{\alpha}$ & $b_{\eta\alpha}$ & $\chiSquareMin/DOF, \,\, \mathrm{probability}$  \\
\noalign{\smallskip}
\hline
\noalign{\smallskip}
$z_{m}<2.48$ & $ 1.052\pm0.055 $ & $ 0.932\pm0.062 $ & $ 2.03\pm0.36 $ & $ -0.233\pm0.017 $ & $ 3192.32 / (3180-~14),  p = 0.37 $ \\
$z_{m}>2.48$ & $ 1.112\pm0.055 $ & $ 0.907\pm0.061 $ & $ 3.20\pm0.79 $ & $ -0.343\pm0.026 $ & $ 3272.11 / (3180-~14),  p = 0.09 $ \\
\noalign{\smallskip}
\hline
\end{tabular}
\end{table*}

\begin{figure*}
\centering
\includegraphics[width=\columnwidth]{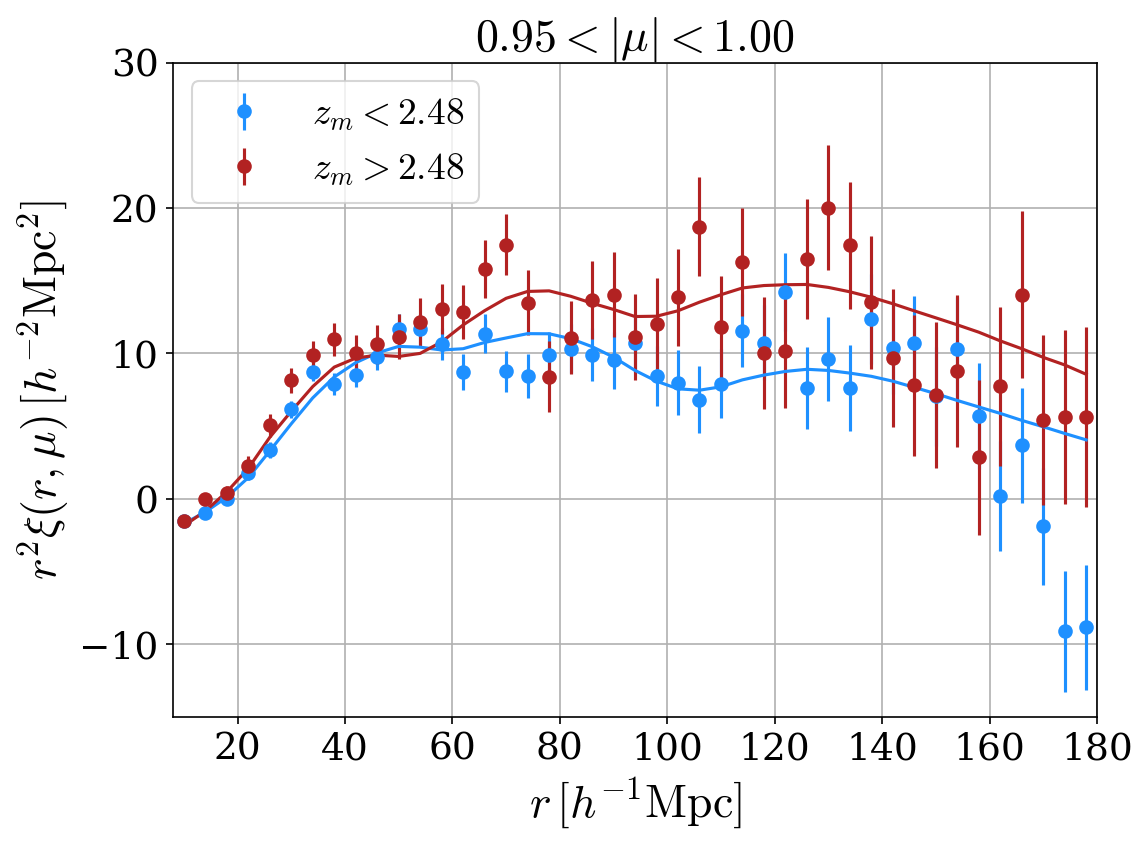}
\includegraphics[width=\columnwidth]{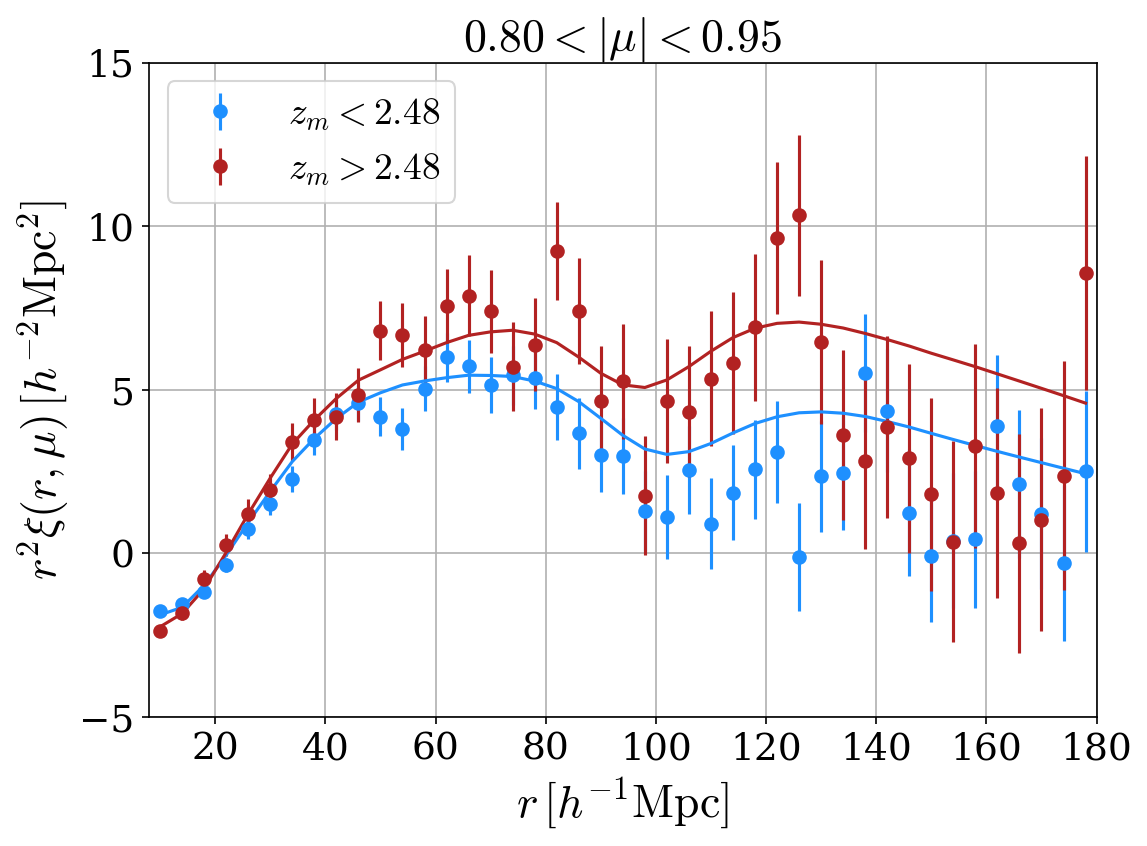}
\includegraphics[width=\columnwidth]{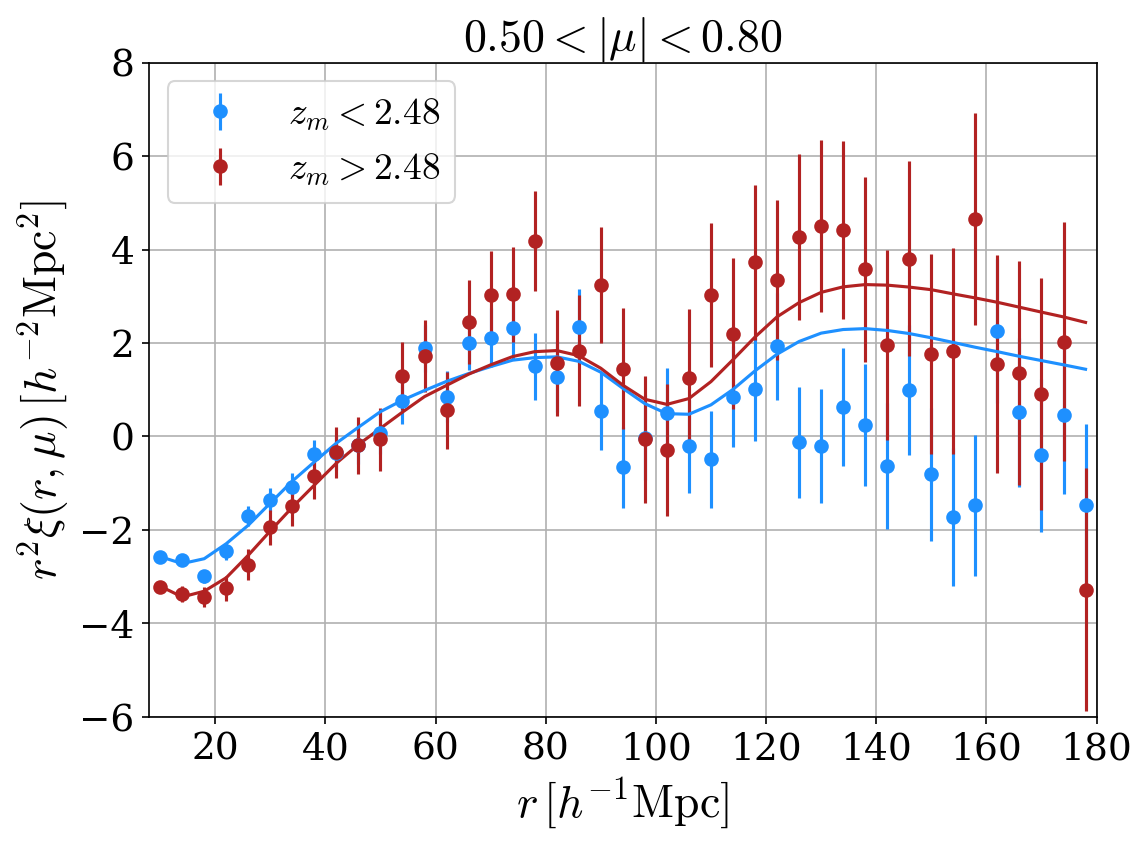}
\includegraphics[width=\columnwidth]{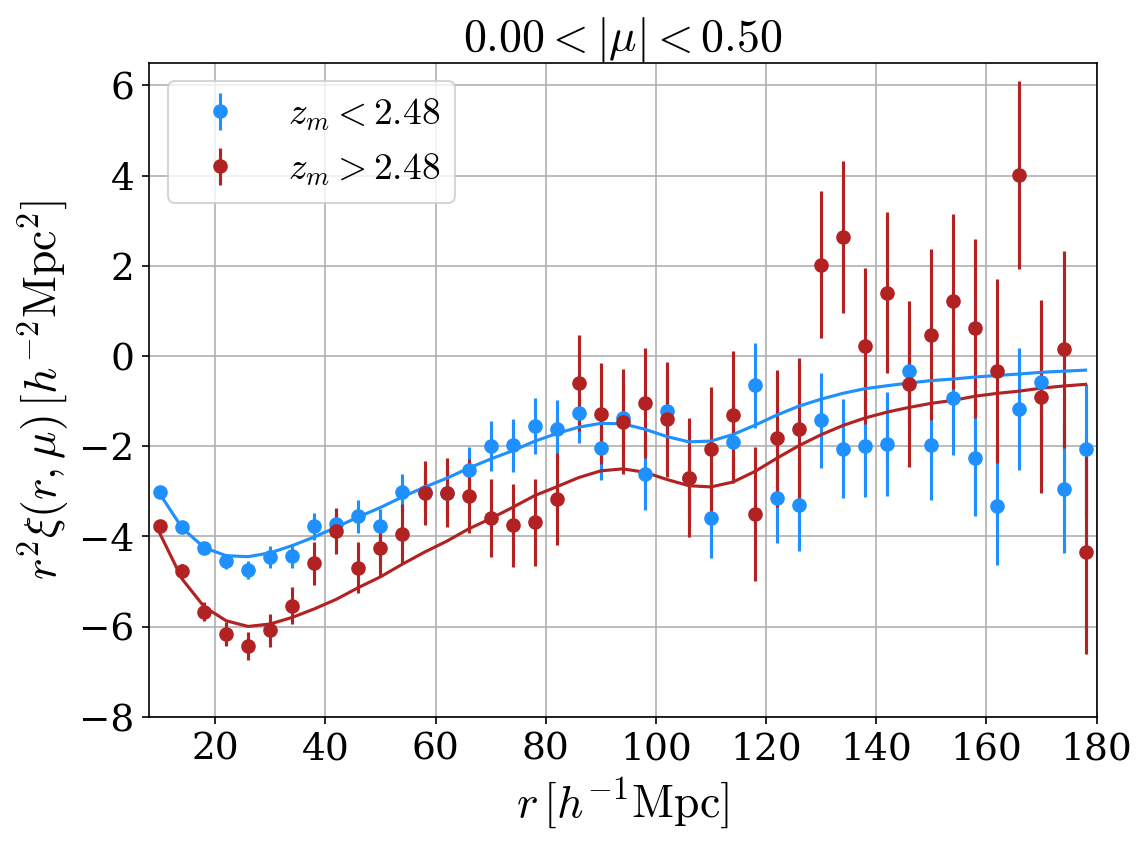}
\caption{Cross-correlation function averaged in four ranges of $\mu=r_{\parallel}/r$ for the fitting range $10<r<180~\hMpc$. The blue points are the data for the low-$z$ bin ($z_{m}<2.48$) and the blue curve the best-fit model. The red points are the data for the high-$z$ bin ($z_{m}>2.48$) and the red curve the best-fit model.}
\label{figure::xi_zbins_wedges}
\end{figure*}

The statistical limitations of the present data set are such that it is not possible to usefully measure the expected redshift-variation of $\DMm(z)/r_d$ and $\DHh(z)/r_d$. However, to search for unexpected effects, we perform an analysis that independently treats a low- and a high-redshift bin.

A quasar and entire forest pair is assigned to either bin depending on their mean redshift:
\begin{equation}
z_{m} = \frac{z_{i,{\rm max}}+z_{\rm q}}{2}\ ,
\end{equation}
where $z_{i,{\rm max}}$ is the pixel with the highest absorption redshift in the forest. As the data split is defined, individual forests and quasars can contribute to both redshift bins. The limiting value of $z_{m}$ is chosen so as to approximately equalize the correlation signal-to-noise ratio (as determined by the best-fit fiducial correlation model) on BAO scales for the two redshift bins. This approach ensures that the redshift bins have similar statistical power for determining the BAO peak position. We set the limit at $z_{m}=2.48$. After identifying which quasar-forest pairs contribute to each redshift bin, we rederive the delta fields for each bin separately to ensure that the mean deltas vanish. The effective redshifts are $z_{\rm eff}=2.21$ and $z_{\rm eff}=2.58$ for the low-$z$ and high-$z$ bin, respectively. The pair redshift distribution for the low-$z$ bin extends up to $z=2.48$ (by definition) and its overlap with the distribution for the  high-$z$ bin is $\Delta z\approx0.25$. Correlations between the redshift bins are at the per cent level.

The result of the data split is summarized in Table~\ref{table::zbins}. Figure~\ref{figure::xi_zbins_wedges} shows the correlation functions and the best-fit models for four ranges of $\mu$. The best-fit values of $(\aperp,\apar)$ for the two bins are consistent, with similar BAO errors of $\sim6\%$. The bias parameter $b_{\eta\alpha}$ changes between the two redshifts by a factor $1.57\pm0.15$ consistent with the expected factor $(3.58/3.21)^{2.9}=1.37$. The parameter $\beta_\alpha$ increases by a factor $1.6\pm0.4$, within two standard deviations of the predicted decrease of 6\% from simulations of \citet{2015JCAP...12..017A}.

\end{document}